\documentclass[11pt]{article}
\usepackage{amssymb}

\usepackage{a4}
\usepackage{vmargin}
\usepackage[fleqn]{amsmath}
\usepackage{euscript}
\usepackage{cite}
\setmarginsrb{3.2cm}{2.5cm}{3.2cm}{2.5cm}{1cm}{1cm}{1cm}{1.6cm}

\makeatletter
\@addtoreset{equation}{section}
\makeatother
\renewcommand{\theequation}{\thesection.\arabic{equation}}
\newcommand{\be}[1]{\begin{equation}\label{#1}}
\newcommand{\bal}{\begin{align}}
\newcommand{\bmult}[1]{\begin{multline}\label{#1}}
\newcommand{\ee}{\end{equation}}
\newcommand{\eal}{\end{align}}

\newtheorem{thm}{\sc Theorem}[section]
\newtheorem{prop}{\sc Proposition}[section]
\newtheorem{cor}{\sc Corollary}[section]
\newtheorem{lemma}{\sc Lemma}[section]
\newtheorem{rem}{Remark}[section]

\def\Proof{\medskip\noindent {\it Proof --- \ }}

\let\qed=\cqfd

\newcommand{\bra}[1]{\langle\,#1\,|}
\newcommand{\ket}[1]{|\,#1\,\rangle}
\newcommand{\bracket}[1]{\langle\,#1\,\rangle}

\def\tr{\operatorname{tr}}

\def\End{\operatorname{End}}

\def\sh{\sinh}
\def\ch{\cosh}

\def\sgz{\sigma^z}
\def\pour#1{_{\,\vrule height 13pt depth 1pt\> {#1}\!}}
\def\sul{\sum\limits}
\def\pl{\prod\limits}

\def\tens{\mathop\otimes\limits}

\def\build#1_#2^#3{\mathrel{
\mathop{\kern 0pt#1}\limits_{#2}^{#3}}}

\newcommand{\Rset }{{\mathbb R }}

\newcommand{\Cset }{{\mathbb C}}

\def\cH{\mathcal{H}}
\def\cA{\mathcal{A}}
\def\cB{\mathcal{B}}
\def\cC{\mathcal{C}}
\def\cD{\mathcal{D}}
\def\cU{\mathcal{U}}
\def\cT{\mathcal{T}}

\def\cS{\mathcal{S}}

\newcommand{\eT}{{\EuScript{T}}}
\newcommand{\eH}{{\EuScript{H}}}
\newcommand{\eV}{{\EuScript{V}}}

\def\eps{\varepsilon}

\def\sg{\sigma}
\def\la{\lambda}

\def\tend{\rightarrow}
\def\ttend{\longrightarrow}

\newcommand{\negspace}{\!\!\!\!}

\newcommand\beq{\begin{equation}}
\newcommand\enq{\end{equation}}

\def\beqa{\begin{eqnarray}}
\def\eeqa{\end{eqnarray}}
\def\ba{\begin{array}}
\def\ea{\end{array}}

\def\a{\alpha}

\def\lt({\left(}
\def\rt){\right)}

\newcommand{\f}[2]{{\ensuremath{%
    \mathchoice%
    {\dfrac{#1}{#2}}
    {\dfrac{#1}{#2}}
    {\frac{#1}{#2}}
    {\frac{#1}{#2}}
}}}
\newcommand{\tf}[2]{\ensuremath{#1/#2}}
\newcommand{\pa}[1]{\ensuremath{\left(#1\right)}}
\newcommand{\paa}[1]{\ensuremath{\left\{#1\right\}}}
\newcommand{\pac}[1]{\ensuremath{\left[#1\right]}}
\newcommand{\paf}[2]{\ensuremath{\left(\f{#1}{#2}\right)}}

\newcommand{\mc}[1]{\ensuremath{\mathcal{#1}}}

\newcommand{\dd}{\text{d}}

\newcommand{\intn}[2]{\ensuremath{[\![ \, #1 \,;\, #2 \,]\!]}}

\begin{document}

\title{{\bf Correlation functions of the open XXZ chain I}}

\author{N.~Kitanine\footnote{LPTM, Universit\'e de Cergy-Pontoise et CNRS, France,
kitanine@ptm.u-cergy.fr},~~
K.~K.~Kozlowski\footnote{ Laboratoire de Physique, ENS Lyon et CNRS,
France,
 karol.kozlowski@ens-lyon.fr},~~
J.~M.~Maillet\footnote{ Laboratoire de Physique, ENS Lyon et CNRS,
France,
 maillet@ens-lyon.fr},\\
G.~Niccoli\footnote{LPTM, Universit\'e de Cergy-Pontoise et CNRS,
France, Giuliano.Niccoli@ens-lyon.fr},~~
N.~A.~Slavnov\footnote{ Steklov Mathematical Institute, Moscow, Russia, nslavnov@mi.ras.ru},~~
V.~Terras\footnote{ Laboratoire de Physique, ENS Lyon et CNRS,
France, veronique.terras@ens-lyon.fr, On leave of absence from LPTA,
Universit\'e Montpellier II et CNRS, France} }


\begin{flushright}
LPENSL-TH-07\\
\end{flushright}
\par \vskip .1in \noindent

\vspace{24pt}

\begin{center}
\begin{LARGE}
{\bf   Correlation functions of the open XXZ chain I}
\end{LARGE}

\vspace{50pt}

\begin{large}
{\bf N.~Kitanine}\footnote[1]{LPTM, Universit\'e de Cergy-Pontoise
et CNRS, France, kitanine@ptm.u-cergy.fr},~~
{\bf K.~K.~Kozlowski}\footnote[2]{ Laboratoire de Physique,  ENS
Lyon et CNRS,  France,
 karol.kozlowski@ens-lyon.fr},~~
{\bf J.~M.~Maillet}\footnote[3]{ Laboratoire de Physique, ENS Lyon
et CNRS, France,
 maillet@ens-lyon.fr},\\
\vspace{0.1cm}
{\bf G.~Niccoli}\footnote[4]{LPTM, Universit\'e de Cergy-Pontoise et
CNRS, France, Giuliano.Niccoli@ens-lyon.fr},~~
{\bf N.~A.~Slavnov}\footnote[5]{ Steklov Mathematical Institute, Moscow, Russia, nslavnov@mi.ras.ru},~~
{\bf V.~Terras}\footnote[6]{ Laboratoire de Physique, ENS Lyon et
CNRS, France, veronique.terras@ens-lyon.fr, on leave of absence from
LPTA, Universit\'e Montpellier II et CNRS, France.} \par

\end{large}
\end{center}

\vspace{3cm}

\begin{abstract}
We consider the $XXZ$ spin chain with diagonal boundary conditions
in the framework of algebraic Bethe Ansatz. Using the explicit
computation of the scalar products of Bethe states and a revisited
version of the bulk inverse problem, we calculate the elementary
building blocks for the correlation functions. In the limit of
half-infinite chain, they are obtained as multiple integrals of
usual functions, similar to the case of periodic boundary
conditions.
\end{abstract}

\newpage

\tableofcontents

\newpage

\section{Introduction}
\label{sec-intro}

Doped low-dimensional antiferromagnets have attracted a lot of
studies especially since the discovery of high-Tc superconductivity.
A particularly simple form of doping results from replacing some
magnetic ions of the crystal by nonmagnetic one's. Open Heisenberg
 quantum spin chains \cite{Hei28} are the archetype of
one-dimensional models providing microscopic description of such
systems. Indeed, the presence of non-magnetic impurities into
crystals having effective one-dimensional magnetic behavior has
drastic effects on their low energy properties : the chain is cut
into finite pieces with essentially free (open) boundaries leading
to the breaking of translational invariance. As a consequence,
physical quantities such as for example the magnetic susceptibility
will get measurable corrections due to the presence of the boundary
\cite{Fuj03,FujE04,FurH04,HikF04,BorS05,GohBF05,SirB06,SirB06a,SirLFEA07,ChaKBDMEAY04
}. The same quantum spin chains have also acquired recently an
important role in the study and the understanding of the interplay
between quantum entanglement and quantum criticality
\cite{VidLRK03,CalC04,CalC05,OstAFF02,OsbN02,RefM04,Kor04,Laf05,LafSCA06}.
There, the presence of boundaries also leads to noticeable effects,
like in particular Friedel oscillations \cite{EggG95,FabG95,BedBFN}
and the algebraic decrease of the boundary part of the entanglement
entropy as a
function of the distance to the boundary \cite{Laf05,LafSCA06}.\\

Correlation functions are central in the description of such effects
and are in fact accessible in experiments.  In particular the local
magnetic susceptibility in the presence of boundary can be obtained
using muon spin rotation/relaxation on the corresponding crystals,
see for example \cite{ChaKBDMEAY04}. More generally, correlation
functions contain the necessary information to compare the
microscopic models at hand to the reality, in particular through the
measurements of dynamical structure factors accessible by neutron
scattering experiments
\cite{Bloch1936,Schwinger1937,Halpern1938,VanHove1954,VanHove1954a,Balescu1975,MarshallL1971}.
While the computation of exact spectrum of Heisenberg chains has
already a very long history, see e.g.
\cite{Bet31,Hul38,Orb58,Wal59,YanY66,YanY66a,LieM66L,FadST79,FadT79,Tha81,Bax82L,Gau83L,BogIK93L,JimM95L},
and references therein, computation of exact correlation functions
of integrable lattice models such as Heisenberg spin chains, in
particular out of their free fermion point where already
considerable work was necessary
\cite{Ons44,Kau49,KauO49,Yan52,LieSM61,McC68,WuMTB76,McCTW77a,SatMJ78m},
has been a major challenge for the last twenty years. Progress have
been obtained using different routes and several results are now
available for the correlation in the bulk, i.e., far from the
boundaries
\cite{IzeK84,IzeK85,BogIK93L,JimMMN92,JimM95L,JimM96,MaiS00,KitMT99,KitMT00,MaiT00,KitMST02a,KitMST02b,KitMST02c,
KitMST02d,KitMST04a,KitMST04b,KitMST04c,GohKS04,GohKS05,GohKS05a,GohHS05,KitMST05a,KitMST05b,CauM05,CauHM05,PerSCHMWA06,
BooJMST06,BooJMST06a,BooJMST06b,BooJMST07}, although still more
progress is needed to obtain full answers. Advances have been
obtained also in the presence of a boundary using in particular
$q$-vertex operator methods \cite{JimKKKM95,JimKKMW95} and field
theory approach \cite{Aff88,AffL91,EggA92,EggA95,WesH00,BilW99,Bil00,Bil00a,WhiAS02,BorS05,GohBF05,SirB06,SirB06a,SirLFEA07}.\\

The aim of the present paper is to develop a method to compute
correlation functions of integrable open (finite and semi-infinite)
spin chains in the framework of the (algebraic) Bethe ansatz for
boundary integrable models
\cite{Gau71,AlcBBBQ87,Skl88,Che84,KulS91,MezN91,KulS92,GhoZ94,FenS94,SkoS95,LecMSS95,KapS96,LesSS96}.
For this purpose, we will consider the example of the finite $XXZ$
spin-1/2 Heisenberg chain with diagonal boundary conditions,
including in particular non-zero boundary longitudinal magnetic
fields, and its (semi-infinite) thermodynamic limit. Our results
concern the general elementary blocks of correlation functions at
zero temperature, namely the average value of arbitrary products
spin operators going from the boundary to an arbitrary site at
distance $m$ from this boundary. Any correlation function can be
written in terms of these elementary blocks. Previous attempts
towards this goal in the Bethe ansatz framework can be found in
\cite{Wan00,Wan02,Wan03,GohBF05}.

The strategy we will follow to solve this problem is closely related
to the one used in the periodic case \cite{KitMT99,KitMT00}. The
central object in this approach is provided by the monodromy matrix
of the open chain, which is a function of a complex spectral
parameter $\lambda$ and of inhomogeneous parameters $\xi_i$ attached
to each site of the chain. Following Sklyanin \cite{Skl88}, it is
given as a quadratic expression in terms of the standard (bulk)
monodromy matrix with the adjunction of the so called boundary
$K$-matrix which encode the boundary conditions
\cite{Skl88,Che84,GhoZ94}; in this paper only diagonal $K$-matrices
will be considered. This boundary monodromy matrix satisfies a
boundary Yang-Baxter algebra governed by two $R$-matrices while the
$K$-matrix itself satisfies its c-number version also called
reflection equation \cite{Skl88}. These settings have been used by
Sklyanin to extend the algebraic Bethe ansatz method to this open
case. In particular, the Hamiltonian of the chain can then be
reconstructed in terms of a weighted (with the $K$ matrix) trace of
this monodromy matrix. Hence as in the periodic case, one can
consider a common set of eigenstates of the boundary transfer matrix
and of the Hamiltonian.

The first task towards the computation of the correlation functions
is to identify the space of states of the open chain as generated by
the action of the entries of the boundary monodromy matrix
(depending on different spectral parameters $\lambda_j$) on some
reference state (here the state with all spins up or down); then
eigenstates of the open chain are obtained from such actions thanks
to the Bethe ansatz equations for the spectral parameters
$\lambda_j$ \cite{Skl88,AlcBBBQ87}. Using this framework, we will
show that it is possible to find determinant expressions for the
scalar product between a boundary Bethe state and an arbitrary
boundary state and consequently  for the norm of the Bethe
eigenstates. This is achieved along the lines used for the bulk case
in \cite{KitMT99} using the factorizing $F$-matrix basis
\cite{MaiS00}.

The second problem is to obtain the action of the local spin
operators on such states. In the bulk case, it was given by the
resolution of the quantum inverse scattering problem, namely, by the
reconstruction of such local operators in site $j$ in terms of a
simple monodromy matrix elements (evaluated at $\lambda = \xi_j$)
multiplied from the right and from the left by products of the
transfer matrices evaluated in the inhomogeneity parameters $\xi_i$
for $i = 1,...,j$. In this bulk case, the Bethe eigenstates of the
Heisenberg chain Hamiltonian being also  common eigenstates for the
transfer matrix, it was straightforward to obtain the explicit
action of the local spin operators on such Bethe states.

The situation in the presence of boundaries turns out to be slightly
more subtle : due to the breaking of translation invariance,
boundary Bethe states are no longer eigenstates of the bulk transfer
matrix, hence leading to a difficult combinatorial problem while
using the
expression of local operators described above.\\
We solve this problem in three steps :\\
(i) We first find a general (simple) relation relating boundary
Bethe states to bulk one's.\\
(ii) Then the reconstruction of local spin operators is obtained
through a rewriting of the above quantum inverse scattering problem
solution as a unique monomial in terms of the bulk monodromy matrix
entries, avoiding in particular the presence of products of transfer
matrices; it gives a new form for the general solution of the
quantum inverse scattering problem.\\
(iii) Due to this new form of the solution of the quantum inverse
scattering problem, the action of the local spin operators on any
boundary Bethe state (expressed in terms of bulk one's) can then be
given in a simple way and the result can be rewritten back in terms
of sums of boundary
Bethe states.\\

Then, using scalar product and norm formulas for boundary states, we
obtain any correlation functions as explicit sums of ratio of
determinants of size half the length of the chain. In the
thermodynamic limit (the limit of semi-infinite chain), these sums
become multiple integrals with weights given in terms of the density
of Bethe roots in the boundary ground state; this density function
indeed describes the infinite size limit of the above ratios of
determinants. For the so-called elementary blocks of correlation
functions it gives proofs of the multiple integrals representations
obtained previously \cite{JimKKKM95,JimKKMW95} using the $q$-vertex
operator method, here both in the massive and massless regimes of
the chain. The problem of computing physical spin correlation
functions will be addressed in a subsequent paper; this involves
summing large number of the elementary blocks obtained here, using
techniques similar to the one's developed for the bulk case
\cite{KitMST02a,KitMST02b}.

This paper is organized as follows. In Section~\ref{sec-notations},
we briefly describe the open $XXZ$ chain with integrable diagonal
boundary conditions and introduce the main notations.
Section~\ref{sec-states} contains some elementary algebraic
properties of boundary operators and boundary states, and a
description of the ground state in the thermodynamic limit. In
Section~\ref{sec-sc-product}, the scalar products of Bethe
eigenstates with arbitrary dual states are computed. In
Section~\ref{sec-act-state}, we explain how, using a new version of
the bulk inverse problem, one can derive the action of a product of
elementary matrices on a boundary arbitrary state. Finally, in
Section~\ref{sec-blocs}, elementary building blocks of the
correlation functions are computed using the results of
Section~\ref{sec-act-state} and Section~\ref{sec-sc-product}, and we
give their multiple integral representation in the thermodynamic
limit. Some technical details are gathered in a set of appendices.

\section{The boundary $XXZ$ chain: definitions and notations}
\label{sec-notations}

In this paper, we consider the $XXZ$ Heisenberg spin-1/2 finite
chain with diagonal boundary conditions. The Hamiltonian of a chain
of $M$ sites is given by
\begin{equation}\label{Ham}
  \cH= \sum_{m=1}^{M-1} \Big\{ \sigma^x_m \,\sigma^x_{m+1} +
  \sigma^y_m\,\sigma^y_{m+1} + \Delta\,(\sigma^z_m\,\sigma^z_{m+1}-1)\Big\}
  + h_-\, \sgz_1 +h_+\,\sgz_M.
\end{equation}
The local spin operators $\sigma^x_m$, $\sigma^y_m$ and $\sigma^z_m$ at site $m$ act as the corresponding Pauli matrices
in the local quantum space ${\eH}_m\sim \Cset^2$, and as the identity operator elsewhere.
The quantum space of states of the chain is ${\eH}={\otimes}_{m=1}^M {\eH}_m$.
In \eqref{Ham}, $\Delta$ is the bulk anisotropy parameter, and $h_\pm$ denote the boundary fields. In what follows, they
will be parametrized as $\Delta=\ch\eta$ and $h_\pm=\sh\eta\,\coth{\xi_\pm} $.

To diagonalize the boundary Hamiltonian $\cH$, we use the modified version of the algebraic Bethe Ansatz proposed by
Sklyanin in \cite{Skl88}. As in the case of periodic boundary conditions, the eigenvectors are obtained as those of a
family of commuting transfer matrices, which are constructed as follows.

Let $R: \Cset \tend\End(V\tens V)$, $V\sim\Cset^2$, denote the
$R$-matrix of the $XXZ$ model,
\be{Rmatrix}
   R(u)= \sh(u+\eta)\,\widehat{R}(u),\qquad \text{with}\quad
   \widehat{R}(u)=
  \begin{pmatrix}
        1&0&0&0 \\
        0&b(u)&c(u)&0 \\
        0&c(u)&b(u)&0 \\
        0&0&0&1
  \end{pmatrix},
\ee
in which
\begin{equation}
  b(u)=\frac{\sinh u}{\sinh (u +\eta )},\quad
  c(u)=\frac{\sinh \eta }{\sinh (u +\eta )}.
\end{equation}
It is obtained as the trigonometric solution of the Yang-Baxter equation\footnote{Here and in the following, indices
label the spaces of the tensor product in which the corresponding operator acts non trivially.
For example, in \eqref{YB}, which is an equation on $V_1\tens V_2\tens V_3$, $V_i\sim\Cset^2$, $R_{ij}$ denotes the
$R$-matrix \eqref{Rmatrix} acting in $V_i\tens V_j$.},
\be{YB}
  R_{12}(u-v)\, R_{13}(u-w)\, R_{23}(v-w)=
  R_{23}(v-w)\, R_{13}(u-w)\, R_{12}(u-v).
\ee
The $R$-matrix satisfies the following initial, unitarity and crossing symmetry relations:
\begin{align}
\label{initial}
 &\widehat{R}(0)= \mathcal{P},\\
\label{unitarity}
 &\widehat{R}_{12}(u)\, \widehat{R}_{21}(-u)= 1,\\
\label{crossing}
 &\sg^y_1\, R^{t_1}_{12}(u-\eta)\, \sg^y_1 = - R_{21}(-u).
\end{align}
Here  $\mathcal{P}$ is the permutation operator on $V\tens V$,
$R_{21}=\mathcal{P}_{12}\, R_{12} \,\mathcal{P}_{12}$, and ${}^{t_1}$ denotes the matrix transposition on the first space
of the tensor product.

Let also $K(u;\xi)$ be the boundary matrix
\be{K}
   K(u)=K(u;\xi)=
         \begin{pmatrix}
             \sh(u+\xi) & 0\\
         0 & \sh(\xi - u)
         \end{pmatrix},
\ee
corresponding to the diagonal solution of the boundary Yang-Baxter equation~\cite{Che84}
\be{bYB}
  R_{12}(u-v)\,K_1(u)\,R_{12}(u+v)\,K_2(v)
  =K_2(v)\,R_{12}(u+v)\,K_1(u)\,R_{12}(u-v).
\ee

A commuting family of transfer matrices $\cT(\la)\in\End\eH$ is constructed from $R$ and $K$ as
\be{transfer}
  \cT(\la) = \tr_0 \{ K_+(\la)\, T(\la)\, K_-(\la)\, \widehat{T}(\la)\}.
\ee
Here the trace is taken over an auxiliary space
$V_0\sim\Cset^2$, $K_\pm (\la)=K(u\pm\eta/2;\xi_\pm)\in \End V_0$,
$T(\la)\in\End(V_0\tens\eH)$ is the bulk monodromy matrix,
\be{T}
  T(\la)=R_{0M}(\la-\xi_M)\ldots R_{02}(\la-\xi_2)\,R_{01}(\la-\xi_1),
\ee
and $\widehat{T}(\lambda)$ is defined as
\be{That}
  \widehat{T}(\la)=R_{10}(\la+\xi_1-\eta)\,R_{20}(\la+\xi_2-\eta)\ldots
       R_{M0}(\la+\xi_M-\eta).
\ee
In these last expressions, $R_{0m}$ denotes the $R$-matrix in $\End(V_0\tens\eH_m)$, and $\xi_1,\ \xi_2,\ldots,\xi_M$ are arbitrary
complex parameters (inhomogeneity parameters) attached to the different sites of the chain.
Note that, due to \eqref{unitarity} and \eqref{crossing},
\begin{align}
  \widehat{T}(\la) &=\gamma(\la)\,
                       \sg^y_0\, T^{t_0}(-\la)\,\sg^y_0 \label{That-T}\\
           &=\widehat\gamma(\la)\,T^{-1}(-\la+\eta),
\end{align}
with, in our normalization,
\begin{equation}
   \gamma(\la)=(-1)^M,\qquad \widehat\gamma(\la)=(-1)^M \prod_{j=1}^M\big[
                          \sh(\la+\xi_j)\,\sh(\la+\xi_j-2\eta)\big].
\end{equation}
In the homogeneous limit ($\xi_m=\eta/2$ for $m=1,\ldots,M$), the Hamiltonian \eqref{Ham} can be obtained as the following
derivative of the transfer matrix \eqref{transfer}:
\be{Ht}
  \cH=\frac{2\,[\sh\eta]^{1-2M}}{\tr\{ K_+(\eta/2)\}\,\tr\{ K_-(\eta/2)\}}
  \frac{d}{d\la}\cT(\la)\pour{\la=\eta/2}+\text{constant}.
\ee

In the case of periodic boundary conditions, the space of states is constructed in terms
of the operator entries $A$, $B$, $C$, $D\in\End\eH$ of the bulk monodromy
matrix \eqref{T} expressed as a $2\times 2$ matrix acting on the auxiliary space:
\be{matrixT}
  T(\la)=\begin{pmatrix}
       A(\la) & B(\la)\\
       C(\la) & D(\la)
     \end{pmatrix}.
\ee
These operators satisfy a quadratic algebra given by the following quadratic relation on $V_1\tens V_2\tens \eH$,
$V_i\sim\Cset^2$:
\be{RTT}
  R_{12}(\la-\mu)\, T_1(\la)\, T_2(\mu) = T_2(\mu)\, T_1(\la)\, R_{12}(\la-\mu).
\ee
In this framework, eigenstates of the periodic Hamiltonian are constructed as the multiple action of creation
operators $B(\la_j)$ on the reference state $\ket{0}$ with all spins up,
provided that the corresponding spectral parameters $\la_j$ satisfy the bulk Bethe equations.

In the case of the diagonal boundary conditions \eqref{Ham},
a similar construction can
be performed (see \cite{Skl88}) using the operators entries
$\cA_-$, $\cB_-$, $\cC_-$, $\cD_-\in\End\eH$ (respectively $\cA_+$, $\cB_+$, $\cC_+$, $\cD_+$)
of one of the ``double-row'' monodromy matrices $\cU_-$
or $\cU_+$ defined on $\End(V_0\tens\eH)$ as
\begin{align}
  \label{U-}
  &\cU_-(\la)
            =T(\la)\, K_-(\la)\, \widehat{T}(\la)
            =\begin{pmatrix}
          \cA_-(\la) & \cB_-(\la)\\
          \cC_-(\la) & \cD_-(\la)
         \end{pmatrix},\\
  \label{U+}
  &\cU_+^{t_0}(\la)
            =T^{t_0}(\la)\, K_+^{t_0}(\la)\, \widehat{T}^{t_0}(\la)
            =\begin{pmatrix}
          \cA_+(\la) & \cC_+(\la)\\
          \cB_+(\la) & \cD_+(\la)
         \end{pmatrix}.
\end{align}
Note that the matrix $\cU_-$ (respectively $\cU_+$), as well as its
operator entries $\cA_-$, $\cB_-$, $\cC_-$ and $\cD_-$ (respectively
$\cA_+$, $\cB_+$, $\cC_+$ and $\cD_+$) depend also on the parameters
$\xi_1,\ldots,\xi_M$ and $\xi_-$ (respectively $\xi_1,\ldots,\xi_M$
and $\xi_+$). It will be sometimes necessary in the paper to specify
explicitly this dependency, denoting e.g. $\cU_-(\la;\xi_-)$ instead
of $\cU_-(\la)$. The matrices $\cU_-$ and $\cU_+$ satisfy the
boundary Yang-Baxter equations
\begin{align}
  &R_{12}(u-v)\,(\cU_{-})_1(u)\,R_{12}(u+v-\eta)\,(\cU_{-})_2(v)\nonumber\\
  &\hspace{3.2cm}
  =(\cU_{-})_2(v)\,R_{12}(u+v-\eta)\,(\cU_{-})_1(u)\,R_{12}(u-v),
             \label{bYB-}\\
  &R_{12}(-u+v)\,(\cU_{+})_1^{t_1}(u)\,
     R_{12}(-u-v-\eta)\,(\cU_{+})_2^{t_2}(v)\nonumber\\
  &\hspace{3.2cm}
  =(\cU_{+})_2^{t_2}(v)\,R_{12}(-u-v-\eta)\,(\cU_{+})_1^{t_1}(u)\,R_{12}(-u+v),
             \label{bYB+}
\end{align}
which leads to commutation relations for their operator entries.

Note that the transfer matrices \eqref{transfer} can be
expressed either in terms of the matrix elements of $\cU_-$,
\begin{align}
  \cT(\la) &= \tr_0 \{ K_+(\la)\,\cU_-(\la)\} \nonumber\\
           &= \sh(\la+\eta/2+\xi_+)\,\cA_-(\la)
                - \sh(\la+\eta/2-\xi_+)\,\cD_-(\la),\label{T-U-}
\end{align}
or in terms of the matrix elements of $\cU_+$,
\begin{align}
  \cT(\la) &= \tr_0 \{ K_-(\la)\,\cU_+(\la)\} \nonumber\\
           &= \sh(\la-\eta/2+\xi_-)\,\cA_+(\la)
                    - \sh(\la-\eta/2-\xi_-)\,\cD_+(\la), \label{T-U+}
\end{align}
and their common eigenstates can be constructed either in the form
\be{BBB-}
  \ket{\psi_-(\{\la\})} = \prod_{k=1}^N \cB_-(\la_j)\ket{0},\qquad
  \bra{\psi_-(\{\la\})} = \bra{0}\prod_{k=1}^N \cC_-(\la_j),
\ee
or in the form
\be{BBB+}
  \ket{\psi_+(\{\la\})} = \prod_{k=1}^N \cB_+(\la_j)\ket{0},\qquad
  \bra{\psi_+(\{\la\})} = \bra{0}\prod_{k=1}^N \cC_+(\la_j),
\ee
provided the set of spectral parameters $\{\la\}$ satisfies the Bethe equations
\begin{equation} \label{eqBethe}
   y_j(\la_j;\{\la\};\xi_+,\xi_-)=y_j(-\la_j;\{\la\};\xi_+,\xi_-),
          \quad j=1,\ldots,N,
\end{equation}
with
\begin{align}
  &y_j(\mu;\{\la\};\xi_+,\xi_-)
       =\frac{\hat{y}(\mu;\{\la\};\xi_+,\xi_-)}
             {\sh(\la_j-\mu+\eta)\,\sh(\la_j+\mu-\eta)},\\
  &\hat{y}(\mu;\{\la\};\xi_+,\xi_-)=-a(\mu)\,d(-\mu)\,\sh(\mu+\xi_+-\eta/2)\,\sh(\mu+\xi_--\eta/2)\nonumber\\
   &\hspace{5cm}\times
      \prod_{k=1}^N \big[\sh(\mu-\la_{k}-\eta) \,\sh(\mu+\la_{k}-\eta)\big].
     \label{fun_y}
\end{align}
Here $a(\la)$ and $d(\la)$ stand respectively for the eigenvalue of the bulk operators $A(\la)$ and $D(\la)$ on the reference
state $\ket{0}$,
\be{ad}
 a(\la)=\prod_{i=1}^M \sh(\la-\xi_i+\eta),\quad
 d(\la)=\prod_{i=1}^M \sh(\la-\xi_i).
\ee
The corresponding eigenvalue of the transfer matrix $\cT(\mu)$ on an eigenstate
\eqref{BBB-} or \eqref{BBB+} is
\begin{multline}
   \tau(\mu, \{\lambda_j\})
     =\gamma(\mu)  \bigg\{  a(\mu) d(-\mu)
   \frac{\sh(2\mu+\eta) \sh(\mu+\xi_+-\eta/2) \sh(\mu+\xi_--\eta/2)}
        {\sh(2\mu)\prod_{i=1}^N [b(\la_i-\mu)\,b(-\mu-\la_i)]} \\
   +  a(-\mu)d(\mu)
   \frac{\sh(2\mu-\eta)\sh(\mu-\xi_+ +\eta/2)\sh(\mu-\xi_- +\eta/2)}
        {\sh(2\mu)\prod_{i=1}^N [b(\mu+\la_i)\,b(\mu-\la_i)]} \bigg\}.
\label{valpropretau}
\end{multline}
Let us finally introduce some convenient notations that we will use all along the paper: for any set of complex
variables $\{x_j\}$, we define
\begin{equation}
x_{jk}=x_{j}-x_{k}\text{ \ \ and \ \ }\overline{x}_{jk}=x_{j}+x_{k}.
\label{short-var}
\end{equation}
%

\section{Boundary states}
\label{sec-states}

\subsection{Algebraic elementary properties}
\label{sec-el-prop}

In this subsection, we collect some usefull elementary properties
concerning boundary operators and boundary states. They mainly
follow from the description of the boundary $XXZ$ model in terms of
the bulk one. Indeed, the ``double-row'' monodromy matrices
$\mathcal{U}_{\pm }$ of the boundary $XXZ$ model being quadratic in
terms of the bulk monodromy matrix $T$ (see definitions
(\ref{U-})-(\ref{U+}) and formula (\ref{That-T})), the boundary
operators are themselves quadratic in terms of the bulk operators.

This quadratic nature influences non-trivially the
dependence on the spectral parameter of the boundary operators; in
particular, a ``$\mathbb{Z}_{2}$ \textit{invariance''} arises in the
spectral parameter dependence of the operators $\mathcal{B}_{\pm }$ and $%
\mathcal{C}_{\pm }$. More precisely, the following proposition holds:

\begin{prop}\label{prop-Sym-B-C}
The boundary operators $\mathcal{B}_{\pm }$ and $\mathcal{C}_{\pm }$ satisfy
the properties:
\begin{alignat*}{2}
  &\mathcal{B}_{-}(-\lambda )
  =-
    \frac{\sinh (2\lambda +\eta )}{\sinh (2\lambda -\eta )}\,
    \mathcal{B}_{-}(\lambda ), \quad &
  &\mathcal{C}_{-}(-\lambda )
  =-
    \frac{\sinh (2\lambda +\eta )}{\sinh (2\lambda -\eta )}\,
    \mathcal{C}_{-}(\lambda ),
         \\
  &\mathcal{B}_{+}(-\lambda )
  =-
    \frac{\sinh (2\lambda -\eta )}{\sinh (2\lambda +\eta )}\,
    \mathcal{B}_{+}(\lambda ), \quad &
  &\mathcal{C}_{+}(-\lambda )
  =-
    \frac{\sinh (2\lambda -\eta )}{\sinh (2\lambda +\eta )}\,
    \mathcal{C}_{+}(\lambda ).
\end{alignat*}
\end{prop}

\textit{Proof --- \ }
Such properties are simple consequences of the
boundary-bulk operator decompositions following from \eqref{U-}-\eqref{U+}.
\qed

Note that the proportionality factors appearing in Proposition \ref{prop-Sym-B-C} are not intrinsic and could in principle be
removed with an appropriate choice of the normalization of the $K$-matrix.

This symmetry has important consequences since the operators
$\mathcal{B}_{-}$ or $\mathcal{B}_{+}$ (respectively
$\mathcal{C}_{-}$ or $\mathcal{C}_{+}$) generate the quantum space
of states of the boundary $XXZ$ model by their multiple action on
the reference state $|\,0\,\rangle $ (respectively, on the dual
reference state $\langle \,0\,|$). In particular, the previous
proposition naturally suggests that the solutions of the Bethe
equations (\ref{eqBethe}) are characterized by the same
$\mathbb{Z}_{2}$ symmetry, which indeed can be shown from a  direct
study of the Bethe equations.

\begin{prop}
\label{prop 4}
Let $\{\lambda _{1},\ldots,\lambda _{N}\}$ be a
solution of the system of Bethe equations \eqref{eqBethe}, then
$\{\sigma _{1}\lambda_{1},\ldots,\sigma _{N}\lambda _{N}\}$ is still a solution for $\sigma _{j}=\pm, $ $j=1,\ldots,N$.
\end{prop}

\textit{Proof --- \ }
This follows directly from the form of the Bethe equations \eqref{eqBethe}.
\qed

As in the bulk case, the operators entries of
the boundary monodromy matrix can be related by some simple relations.
This is the subject of the next lemma.

\begin{lemma}
\label{prop 1}
The following relations hold:
\begin{equation}
\sigma _{0}^{x}\, \cU_{\pm }(\lambda ; \xi_\pm )\, \sigma _{0}^{x}
  =-\Gamma _{x}\, \cU_{\pm }(\lambda ; -\xi_\pm )\,\Gamma_x,
\end{equation}
or explicitly:
\begin{equation}
 \mathcal{A}_{\pm }(\lambda; \xi_\pm )
  =-\Gamma _{x}\,\mathcal{D}_{\pm }(\lambda; -\xi_\pm  )\,\Gamma _{x},
            \qquad
 \mathcal{C}_{\pm }(\lambda; \xi_\pm )
  =-\Gamma _{x}\,\mathcal{B}_{\pm }(\lambda; -\xi_\pm  )\,\Gamma _{x},
\end{equation}
where
$
\Gamma _{x}=\underset{k =1}{\overset{M}{%
\otimes }}\sigma _{k}^{x}.
$
\end{lemma}

\textit{Proof --- \ }
These identities follow from the definitions \eqref{U-}-\eqref{U+} and from the bulk identity
$\sigma_0^x\,T(\la)\,\sigma_0^x=\Gamma _{x}\,T(\lambda )\,\Gamma _{x}$.
\qed

The question now arises whether the state $\bra{\psi_-(\{\la\})}$ \eqref{BBB-}
(respectively $\bra{\psi_+(\{\la\})}$ \eqref{BBB+})
is actually related to the dual state of $\ket{\psi_-(\{\la\})}$ (respectively $\ket{\psi_+(\{\la\})}$), i.e.
whether the operators $\cB_-(\la)$ and $\cC_-(\la)$ (respectively  $\cB_+(\la)$ and $\cC_+(\la)$) are conjugated to each other.
Indeed, if the Hermitian conjugate $V^\dagger$ of an operator $V\in\End(V_0\tens\eH)$ is defined as
\begin{equation}
V^{\dagger }( \lambda) =\big[V(\lambda )\big]^{t_{1}\ldots t_{M}\,\ast },
\end{equation}
where $^{t_{1}\ldots t_{M}}$ denotes the transposition on the quantum space $\eH$ and $^{\ast }$ the
complex conjugation on $c$-numbers, we have the following result.

\begin{prop}
\label{Dual}
In the vicinity of the homogeneous limit of the massless model
($\eta \in i\mathbb{R}$,  $\xi_{k}-\eta/2 \in \mathbb{R}$ and $\xi _{\pm}\in i\mathbb{R}$), $\cU_{\pm }\left( \lambda \right) $
has the following Hermitian
conjugate:
\begin{equation}
  \cU_{\pm }^{\dagger } ( \lambda )
  =- \big\{ \cU_{\pm }( -\lambda ^{\ast }) \big\}^{t_0}.
\end{equation}
An analogous result holds in the vicinity of the homogeneous limit of the
massive model ($\eta \in \mathbb{R}$, $\xi _{k}-\eta/2 \in i\mathbb{R}$ and $\xi _{\pm}-i q_\pm \pi/2 \in \mathbb{R}$
with $q_\pm=0,1$), namely
\begin{equation}
\cU_{\pm }^{\dagger }( \lambda ) = (-1)^{q_\pm} \big\{
\cU_{\pm }( \lambda ^{\ast }) \big\}^{t_0}.
\end{equation}
\end{prop}

\textit{Proof --- \ }
It follows from the conjugation properties for $R$ and $K$:
\begin{equation*}
  R_{0i}^{\dagger }\left( \lambda \right) = - R_{0i}^{t_{0}} ( -\lambda ^{\ast }),
    \qquad
  K( \lambda \pm \eta /2;\xi _{\pm })^{\ast }
       =-K( -\lambda^{\ast }\pm \eta /2;\xi _{\pm } ),
\end{equation*}
in the massless case, and
\begin{equation*}
  R_{0i}^{\dagger }( \lambda ) = R_{0i}^{t_0}(\lambda ^{\ast }),
    \hspace{1.4cm} K( \lambda \pm \eta /2;\xi _{\pm })^{\ast }
           = (-1)^{q_\pm} K(\la^{\ast }\pm \eta /2;\xi _{\pm }),
\end{equation*}
in the massive case.
\qed

In the next proposition, a set of formul\ae\ are derived to express
the states of the boundary $XXZ$ model in terms of those of the
periodic bulk $XXZ$ model.

\begin{prop}
\label{prop 3}
Let $\la_1,\ldots,\la_N$ be arbitrary complex numbers. Then the boundary states $\ket{\psi_\eps(\{\la\})}$
and $\bra{\psi_\eps(\{\la\})}$, $\eps=\pm$, can be expressed in terms of the bulk states as
\begin{align}
 &\ket{\psi_\eps(\{\la\})}
    =\sum_{\sigma_{1},\ldots,\sigma _{N}=\pm }
     H_{(\sigma _{1},\ldots,\sigma _{N})}^{\mathcal{B}_{\eps}}
        (\lambda _{1},\ldots,\lambda _{N};\xi_\eps)\
     \prod_{j=1}^{N} B(\lambda_{j}^\sigma)|\,0\,\rangle ,
   \label{B_b-vector}\\
 &\bra{\psi_\eps(\{\la\})}
   =\sum_{\sigma_{1},\ldots,\sigma _{N}=\pm }
    H_{(\sigma _{1},\ldots,\sigma _{N})}^{\mathcal{C}_{\eps}}
       (\lambda _{1},\ldots,\lambda _{N};\xi_\eps)\
    \langle \,0\,|\prod_{j=1}^{N}C(\lambda _{j}^\sigma),
   \label{C_b-vector}
\end{align}
where
\begin{align}
&H_{(\sigma _{1},\ldots,\sigma _{N})}^{\mathcal{B}_{-}}
        (\lambda _{1},\ldots,\lambda_{N};\xi_-)
     =\prod_{j=1}^{N} \Big[ -\sigma _{j}\,\gamma(\la_j)\,
     a(-\lambda _{j}^\sigma)\,
     \frac{\sinh(2\lambda _{j}-\eta )}{\sinh (2\lambda _{j})}
               \notag \\
 &\hspace{4.2cm} \times
     \sinh (\lambda _{j}^\sigma-\xi _{-}+\eta /2)\Big]
     \prod_{1\leq r<s\leq N}
     \frac{\sinh (\bar\lambda _{rs}^\sigma+\eta )}
          {\sinh (\bar\lambda _{rs}^\sigma)},
             \label{HB-} \\
&H_{(\sigma _{1},\ldots,\sigma _{N})}^{\mathcal{C}_{-}}
        (\lambda _{1},\ldots,\lambda_{N};\xi_-)
    =\prod_{j=1}^{N}\Big[ \sigma _{j}\,\gamma(\la_j)\,
     d(-\lambda _{j}^\sigma)\,
     \frac{\sinh(2\lambda _{j}-\eta )}{\sinh (2\lambda _{j})}
               \notag \\
 &\hspace{4.2cm} \times
     \sinh (\lambda _{j}^\sigma+\xi _{-}-\eta /2)\Big]
     \prod_{1\leq r<s\leq N}
     \frac{\sinh (\bar\lambda _{rs}^\sigma-\eta )}
          {\sinh (\bar\lambda _{rs}^\sigma)},
             \label{HC-}\\
&H_{(\sigma _{1},\ldots,\sigma _{N})}^{\mathcal{C}_{+}}
    (\lambda _{1},\ldots,\lambda_{N};\xi_+)
    =\prod_{j=1}^{N} \Big[ -\sigma _{j}\,\gamma(\la_j)\,
     a(-\lambda _{j}^\sigma)\,
     \frac{\sinh(2\lambda _{j}+\eta )}{\sinh (2\lambda _{j})}
               \notag \\
 &\hspace{4.2cm} \times
     \sinh (\lambda _{j}^\sigma-\xi _{+}+\eta /2)\Big]
     \prod_{1\leq r<s\leq N}
     \frac{\sinh (\bar\lambda _{rs}^\sigma+\eta )}
          {\sinh (\bar\lambda _{rs}^\sigma)},
              \label{HC+} \\
&H_{(\sigma _{1},\ldots,\sigma _{N})}^{\mathcal{B}_{+}}
    (\lambda _{1},\ldots,\lambda_{N};\xi_+)
    =\prod_{j=1}^{N} \Big[ \sigma _{j}\,\gamma(\la_j)\,
     d(-\lambda _{j}^\sigma)\,
     \frac{\sinh(2\lambda _{j}+\eta )}{\sinh (2\lambda _{j})}
               \notag \\
 &\hspace{4.2cm} \times
     \sinh (\lambda _{j}^\sigma+\xi _{+}-\eta /2)\Big]
     \prod_{1\leq r<s\leq N}
     \frac{\sinh (\bar\lambda _{rs}^\sigma-\eta )}
          {\sinh (\bar\lambda _{rs}^\sigma)},
\label{HB+}
\end{align}
in which we have used the notations $\la_j^\sigma=\sigma_j\la_j$ and $\bar{\la}_{jk}^\sigma=\sigma_j\la_j+\sigma_k\la_k$
for  $j,k=1,\ldots,N$.
\end{prop}

\textit{Proof ---} \
Let us show \eqref{C_b-vector} and \eqref{HC-} for the state $\bra{\psi_-(\{\la\})}$ by induction on $N$, the proofs
for $\bra{\psi_+(\{\la\})}$ and $\ket{\psi_\pm(\{\la\})}$ being similar.

For $N=1$, the expression follows from the representation \eqref{Decomp-C-}
of the operator $\mathcal{C}_{-}$ and from the action of $D$ on the dual reference state $\bra{0}$.

Let us now suppose that the decomposition \eqref{C_b-vector}-\eqref{HC-} holds for any set of complex
variables $\{\la_1,\ldots,\la_N\}$. The action of $\mathcal{C}_{-}(\lambda _{N+1})$ on the
state $\bra{\psi(\{\la_1,\ldots,\la_N\})}$  can be computed from \eqref{Decomp-C-} using the expression of
the bulk action of $D(\lambda _{N+1})$ \cite{FadST79}:
\begin{equation}
  \bra{0}\prod_{j=1}^{N}C(\lambda _{j})D(\lambda_{N+1})
  =\sum_{k=1}^{N+1} d(\lambda _k)\,
   \frac{\underset{j=1}{\overset{N}{\prod }}\sinh (\lambda _{kj}+\eta )}
        {\underset{\underset{j\neq k}{j=1}}{\overset{N+1}{\prod }}
                                             \sinh \lambda _{kj}}\,
  \bra{0}\underset{\underset{j\neq k}{j=1}}{\overset{N+1}{\prod }}
                       C(\lambda _{j}).  \label{FCactD}
\end{equation}
In such a sum we can distinguish between the direct term $k=N+1$, and the indirect terms $k<N+1.$
Let us show that
each indirect term does not contribute. The indirect terms corresponding to a given $k<N+1$ are proportional to the states
$\bra{0}\prod_{j=1,j\neq k}^{N}C(\lambda _{j}^\sigma)\,
C(\lambda_{N+1})\, C(-\lambda _{N+1})$ with
coefficients:
\begin{multline}
  \gamma(\la_{N+1})\,\sinh \eta \,
    \frac{\sinh (2\lambda_{N+1}-\eta )}{\sinh (2\lambda _{N+1})}
    \sum_{\sigma _{k},\sigma _{N+1}=\pm }
    \sigma _{N+1}\, d(\sigma _{k}\lambda _{k})
    \frac{\sinh(\lambda _{N+1}^\sigma-\xi _{-}+\eta /2)}
         {\sinh (\lambda_{k}^\sigma-\lambda _{N+1}^\sigma)}  \\
  \times \prod\limits_{\underset{j\neq k}{j=1}}^{N}
    \frac{\sinh (\lambda _{kj}^\sigma+\eta)}
         {\sinh (\lambda _{kj}^\sigma)}\,
    H_{(\sigma_{1},\ldots,\sigma _{N})}^{\mathcal{C}_{-}}
        (\lambda _{1},\ldots,\lambda _{N};\xi_-).
\label{coef-induction}
\end{multline}
There we factorize the following expression:
\begin{multline}
  \frac{\sinh (2\lambda _{N+1}-\eta )}{\sinh (2\lambda _{N+1})}
  \frac{\sinh (2\lambda _{k}-\eta )}{\sinh (2\lambda _{k})}
  \prod\limits_{\underset{j\neq k}{j=1}}^{N}
  \frac{\sinh (\lambda _{k}-\lambda_{j}^\sigma+\eta )
        \sinh (\lambda _{k}+\lambda _{j}^\sigma-\eta )}
       {\sinh(\lambda _k-\lambda _{j}^\sigma)
        \sinh (\lambda _{k}+\lambda_{j}^\sigma)}   \\
  \times \sh\eta\ \gamma(\la_{N+1})\,d(\lambda _{k})\, d(-\lambda _{k})\,
    \hat{H}_{k,(\sigma _{1},\ldots,\sigma _{N})}^{\mathcal{C}_{-}}
             (\lambda _{1},\ldots,\lambda_{N};\xi_-),
\end{multline}
which does not depend on the values of $\sigma _{k}$ and $\sigma _{N+1}$.
Here, $\hat{H}_{k,(\sigma _{1},\ldots,\sigma _{N})}^{\mathcal{C}_{-}}(\lambda
_{1},\ldots,\lambda _{N};\xi_-)$ is a part of
$H_{(\sigma _{1},\ldots,\sigma _{N})}^{\mathcal{C}_{-}}(\lambda _{1},\ldots,\lambda _{N};\xi_-)$
which does not contain $\lambda _{k}$.
The remaining sum in (\ref{coef-induction}) reads as
\begin{equation}
\sum_{\sigma _{k},\sigma _{N+1}=\pm }\sigma_{k}\,\sigma _{N+1}\,
\frac{\sinh (\lambda _{k}^\sigma+\xi _{-}-\eta /2)\,
      \sinh (\lambda _{N+1}^\sigma-\xi _{-}+\eta /2)}
     {\sinh (\lambda _{k}^\sigma-\lambda _{N+1}^\sigma)},
\end{equation}
which is zero.

Thus, only the direct action of \eqref{FCactD} contributes, and it generates (\ref{C_b-vector}).
%
\qed

\begin{rem}
The above proposition implies that, for specific values of the spectral parameters, the corresponding boundary
and bulk states are proportional.
For example, since $d(\xi_i)=a(\xi_i-\eta)=0$, we have,
%
\begin{align}
  &\ket{\psi_+(\{\xi_{i_h}\})}
   = H_{1}^{\mathcal{B}_{+}} \big(\{\xi _{i_{h}}\};\xi_+\big)\,
    \prod_{h=1}^{N}B(\xi _{i_{h}}) \ket{0} ,
              \\
  &\bra{\psi_-(\{\xi_{i_h}\})}
  = H_{1}^{\mathcal{C}_{-}} \big(\{\xi _{i_{h}}\};\xi_-\big)\,
   \bra{0}\prod_{h=1}^{N}C(\xi _{i_{h}}),
               \\
  &\ket{\psi_-(\{\xi_{i_h}-\eta\})}
  = H_{1}^{\mathcal{B}_{-}} \big(\{\xi_{i_h}-\eta\} ;\xi_-\big)\,
    \prod_{h=1}^{N}B(\xi _{i_{h}}-\eta) \ket{0} ,
              \\
  &\bra{\psi_+(\{\xi_{i_h}-\eta\})}
  = H_{1}^{\mathcal{C}_{+}} \big(\{\xi_{i_h}-\eta\};\xi_+\big)\,
   \bra{0}\prod_{h=1}^{N}C(\xi _{i_{h}}-\eta),
\end{align}
in which $\{\xi_{i_1},\ldots,\xi_{i_N}\}$ is a subset of $\{\xi_1,\ldots,\xi_M\}$,
and $H_{1}^{\mathcal{O}_\pm} \big(\{\la\};\xi_\pm\big)$, for $\mathcal{O}=\mathcal{B},\mathcal{C}$, denotes the coefficient
$H_{(1,\ldots,1)}^{\mathcal{O}_\pm} (\la_1,\ldots,\la_N;\xi_\pm\big)$.
\end{rem}

Note that the previous proposition, together with the bulk
decompositions of the boundary operators, allows us in principle to
reformulate the quantum inverse problem for the boundary $XXZ$ in
terms of the periodic bulk one. Indeed, we will use this property,
in Section~\ref{sec-act-state}, to compute the action of a product
of local operators on a boundary state.

Now, let us recall that the two expressions (\ref{T-U-})-(\ref{T-U+}) of the
boundary transfer matrix $\mathcal{T}$ coincide as well as the Bethe
equations derived using the $\ket{\psi _{-}} $ boundary
states (\ref{BBB-})\ or the $\ket{\psi _{+}}$ ones (\ref{BBB+}).
In absence of degeneration, these observations naturally suggest that, for
any solution of the boundary Bethe equations, the corresponding eigenstates
$\ket{\psi _{-}(\{\la\})}$ and $\ket{\psi _{+}(\{\la\})}$ have
to be proportional to each other. Indeed, this holds as shown explicitly in
the next proposition.

\begin{prop}
\label{prop 5}
Let $\{\lambda _{1},\ldots,\lambda _{N}\}$ be a solution of
the system of  Bethe equations\;\eqref{eqBethe}. Then the
corresponding eigenstates generated by $\mathcal{B}_{+}$
and $\mathcal{B}_{-}$ are proportional, as well as those
generated by $\mathcal{C}_{+}$ and $\mathcal{C}_{-}$:
\begin{align}
\prod_{j=1}^{N}\mathcal{B}_{+}(\lambda _{j}) \ket{0}
  =\prod_{j=1}^N\frac{\sh(\eta+2\la_j)}{\sh(\eta-2\la_j)}\;
   G(\{\lambda_{a}\};\xi _{+},\xi _{-})
   \prod_{j=1}^{N}\mathcal{B}_{-}(\lambda_{j}) \ket{0}  ,
                       \\
\bra{0} \prod_{j=1}^{N}\mathcal{C}_{-}(\lambda _{j})
  =\prod_{j=1}^N\frac{\sh(\eta-2\la_j)}{\sh(\eta+2\la_j)}\;
   G(\{\lambda_{a}\};\xi _{-},\xi _{+})\,
   \bra{0}\prod_{j=1}^{N}\mathcal{C}_{+}(\lambda_{j}),
\end{align}
where
\begin{equation}
G(\{\lambda _{a}\};x,y)=\prod_{j=1}^{N}\frac{d(\lambda _{j})}{a(\lambda _{j})}
    \frac{\sinh (\lambda_{j}-x+\eta /2)}{\sinh (\lambda _{j}+y-\eta /2)}
   \prod_{1\leq r<s\leq N}
    \frac{\sinh (\lambda _{r}+\lambda_{s}-\eta )}
         {\sinh (\lambda_{r}+\lambda_{s}+\eta )} .
  \label{DEF-G}
\end{equation}
\end{prop}

\textit{Proof --- \ }
The above identities can be proved using the boundary-bulk decomposition
of Proposition~\ref{prop 3}, by directly showing that the two ratios
\begin{align}
&H_{\mathcal{B}_{+}/\mathcal{B}_{-}}
  =(-1)^N \frac{H_{(\sigma _{1},\ldots,\sigma_{N})}^{\mathcal{B}_{+}}
            (\lambda _{1},\ldots,\lambda _{N};\xi_+)}
    {H_{(\sigma _{1},\ldots,\sigma_{N})}^{\mathcal{B}_{-}}
        (\lambda _{1},\ldots,\lambda _{N};\xi_-)},
     \label{ratio-B/B}\\
&H_{\mathcal{C}_{-}/\mathcal{C}_{+}}
  =(-1)^N \frac{H_{(\sigma _{1},\ldots,\sigma _{N})}^{\mathcal{C}_{-}}
            (\lambda _{1},\ldots,\lambda _{N};\xi_-)}
    {H_{(\sigma _{1},\ldots,\sigma_{N})}^{\mathcal{C}_{+}}
        (\lambda _{1},\ldots,\lambda _{N};\xi_+)},
     \label{ratio-C/C}
\end{align}
do not depend on $\{\sigma _{1},\ldots,\sigma _{N}\}$ and coincide respectively with
\begin{equation*}
 \prod_{j=1}^{N}\frac{\sinh (2\lambda _{j}+\eta )}{\sinh (2\lambda _{j}-\eta )}\,
 G(\{\lambda _{a}\};\xi _{+},\xi _{-})
 \quad\text{and}\quad
 \prod_{j=1}^{N}\frac{\sinh (2\lambda _{j}-\eta )}{\sinh (2\lambda _{j}+\eta )}\,
 G(\{\lambda_{a}\};\xi _{-},\xi _{+}).
\end{equation*}
Let us consider for example the ratio $H_{\mathcal{B}_{+}/\mathcal{B}_{-}}$ \eqref{ratio-B/B}, which reads:
\begin{equation*}
  H_{\mathcal{B}_{+}/\mathcal{B}_{-}}=
  \prod_{j=1}^{N}\frac{\sinh (2\lambda _{j}+\eta )}{\sinh (2\lambda _{j}-\eta )}
  \frac{d(-\lambda_{j}^\sigma)\sinh(\lambda_{j}^\sigma+\xi_{+}-\eta /2)}
       {a(-\lambda_{j}^\sigma)\sinh(\lambda_{j}^\sigma-\xi _{-}+\eta /2)}
  \prod_{1\leq r<s\leq N}
    \frac{\sinh (\bar\lambda^\sigma _{rs}-\eta )}
         {\sinh (\bar\lambda^\sigma_{rs}+\eta )}.
\end{equation*}
The action of the transformation $\sigma _{a}\rightarrow -\sigma _{a}$ on
such a ratio for a given $a\in\{1,\ldots,N\}$ gives:
\begin{equation}
\hat{H}_{a,\mathcal{B}_{+}/\mathcal{B}_{-}}
  \frac{\sinh (2\lambda _{a}+\eta )}{\sinh (2\lambda _{a}-\eta )}
  \frac{d(\lambda _{a}^\sigma)\sinh (\lambda^\sigma_{a}-\xi _{+}+\eta /2))}
       {a(\lambda _{a}^\sigma)\sinh (\lambda^\sigma _{a}+\xi _{-}+\eta /2))}
  \prod_{\substack{ s=1  \\ s\neq a }}^{N}
  \frac{\sinh (\lambda^\sigma _{as}+\eta )}{%
        \sinh (\lambda^\sigma _{as}-\eta )},
\label{ratio-(+-)}
\end{equation}
where $\hat{H}_{a,\mathcal{B}_{+}/\mathcal{B}_{-}}$ is a part of $H_{%
\mathcal{B}_{+}/\mathcal{B}_{-}}$ which does not contain $\lambda _{a}$.
Proposition \ref{prop 4} now implies that
$\{\lambda^\sigma_{1},\ldots,\lambda _{N}^\sigma\}$ is a solution of Bethe equations
if $\{\lambda _{1},\ldots,\lambda _{N}\}$ is a solution.
Therefore, applying the Bethe
equation (\ref{eqBethe}) for $j=a$, we have
\begin{multline}
  \frac{d(\lambda _{a}^\sigma)}{a(\lambda _{a}^\sigma)}
  \frac{\sinh (\lambda _{a}^\sigma-\xi_{+}+\eta /2)}
       {\sinh (\lambda _{a}^\sigma+\xi _{-}-\eta /2)}
  \prod_{\substack{ s=1  \\ s\neq a}}^{N}
    \frac{\sinh(\lambda _{as}^\sigma+\eta )}
         {\sinh (\lambda _{as}^\sigma-\eta )}\\
 =
 \frac{d(-\lambda_{a}^\sigma)}{a(-\lambda _{a}^\sigma)}
 \frac{\sinh (\lambda _{a}^\sigma+\xi _{+}-\eta /2)}
      {\sinh (\lambda _{a}^\sigma-\xi _{-}+\eta /2)}
 \prod_{\substack{ s=1  \\ s\neq a}}^{N}
   \frac{\sinh (\bar\lambda_{as}^\sigma-\eta )}
        {\sinh (\bar\lambda _{as}^\sigma+\eta )},
\end{multline}
so that (\ref{ratio-(+-)}) coincides with  the expression of $H_{\mathcal{B}_{+}/\mathcal{B}_{-}}$.
\qed

\subsection{Description of the ground state}\label{sec-gs}

The Bethe equations \eqref{eqBethe} can be written in the logarithmic form as
\begin{equation}
 2M p(\la_j) + g(\la_j;\xi_+,\xi_-)
  + \sul_{\substack{k=1 \\ k\not=j}}^{N}
  \big[ \theta(\la_{jk}) + \theta(\bar{\la}_{jk}) \big] =2\pi n_j,
 \qquad 1\le j \le N
 \label{eqbethelog}
\end{equation}
where $n_j$ are integers (with $n_j<n_{j+1}$), and
where the momentum $p$, the scattering phase $\theta$ and the boundary contribution $g$ are defined as
\begin{align}
  &p(\la)=\frac{i}{2M}\ln\frac{d(\la)\,a(-\la)}{a(\la)\,d(-\la)},\\
  &\theta(\la)=i\ln\frac{\sh(\eta+\la)}{\sh(\eta-\la)},\\
  &g(\la;\xi_+,\xi_-)=i\ln\frac{\sh(\la-\xi_++\eta/2)\,\sh(\la-\xi_-+\eta/2)}
                               {\sh(\la+\xi_+-\eta/2)\,\sh(\la+\xi_--\eta/2)}.
\end{align}
In the homogeneous limit, the corresponding eigenvalues of the Hamiltonian \mc{H} in the spin $M-N$ sector are
\begin{equation}
E(\{\la\})=\sh \eta \, \big\{\coth \xi_+ +\coth \xi_- \big\}
             +4\sul_{j=1}^N\big\{\cos p(\la_j)-\Delta\big\}.
\end{equation}

In order to characterize the ground state of the half-infinite chain $M\tend\infty$, one should distinguish the two
domains $-1<\Delta\le 1$ (massless regime) and $\Delta>1$ (massive regime), for which we set:
\begin{alignat*}{4}
  &\alpha_j=\la_j,\quad  & &\zeta=i\eta>0,\quad &
  &\xi_-=-i\tilde\xi_-,\quad \text{with}\
  -\frac{\pi}{2}<\tilde\xi_-\le\frac{\pi}{2},\qquad  &
  &\text{for}\ 1<\Delta\le 1,\\
  &\alpha_j=i\la_j,\quad  & &\zeta=-\eta>0,\quad &
  &\xi_-=-\tilde\xi_-+i\delta\frac{\pi}{2},\quad\text{with}\
  \tilde\xi_-\in\Rset, \qquad  &
  &\text{for}\ \Delta>1,
\end{alignat*}
in which $\delta=1$ for $| h_-|<\sh\zeta$ and $\delta=0$ otherwise.
Thus, to a given set of roots $\{\la_j\}$ corresponds a set of variables $\{\alpha_j\}$ given
by the previous change of variables.
Note that two sets of Bethe roots $\{\la_j\}$ and $\{\sigma_j\la_j\}$, where $\sigma_j=\pm$, correspond to the same
Bethe vector. Therefore, we consider only solutions $\{\la_j\}$ such that $\Re(\alpha_j)>0$ or
$\Re(\alpha_j)=0,\ \Im(\alpha_j)<0$.

The ground state of the half-infinite chain $M\tend\infty$ has been studied in \cite{SkoS95}, \cite{KapS96}. It appears
that the nature of the ground state rapidities depends on the value of the boundary field $h_-$.

In the case where $\tilde\xi_-<0$ or $\tilde\xi_->\zeta/2$, the ground state of the Hamiltonian \eqref{Ham} is given
in both regimes by the maximum number $N$ of roots $\la_j$ corresponding to real (positive)
$\alpha_j$ such that $\cos p(\la_j) <\Delta$.
In the thermodynamic limit $M\tend\infty$, these roots $\la_j$
form a dense distribution on an interval $[0,\Lambda]$ of the real or imaginary axis. Their density
\begin{equation}
   \rho(\la_j)=\lim_{M\tend\infty} [M(\la_{j+1}-\la_j)]^{-1}
\end{equation}
satifies the following integral equation:
\begin{equation}\label{Lieb}
\rho(\la)+\int\limits_{0}^{\Lambda}
\big[K(\la-\mu)+K(\la+\mu)\big]\,\rho(\la)\, \dd \la
        =\frac{p'(\la) }{\pi}.
\end{equation}
Here,
\begin{equation}
    K(\la)=-\frac{1}{2\pi} \theta'(\la)
    =\frac{i\sh (2\eta)}{2\pi\,\sh(\la+\eta)\,\sh(\la-\eta)},
\end{equation}
and $\Lambda=+\infty$ in the massless regime, while $\Lambda=-i\pi/2$ in the massive one.
We may extend the definition of $\rho$ as the solution of \eqref{Lieb} on the whole
interval $[-\Lambda,\Lambda]$. It is then easy to see that $\rho(-\la)=\rho(\la)$.
Therefore, $\rho$ satisfies the equation
\begin{equation}\label{Liebbulk}
\rho(\la)+\int\limits_{-\Lambda}^{\Lambda}
K(\la-\mu)\,\rho(\la)\, \dd \la
        =\frac{p'(\la) }{\pi},
\end{equation}
which means that the density of Bethe roots for the ground state of the open chain is twice the corresponding density in
the periodic case.

In the case $0<2\tilde\xi_-<\zeta$, the ground state admits also a root $\check\la$ (corresponding to a complex $\check\alpha$)
which tends to $\eta/2-\xi_-$ with exponentially small corrections in the large $M$ limit.
In that case, the real roots density is still given by \eqref{Liebbulk}.

All the above results are valid in the homogeneous limit. However, for technical convenience, we also introduce a familly of
inhomogeneous
densities $\rho(\la,\xi)$, depending on an additional parameter $\xi$, as solutions of the
integral equation
\begin{equation}\label{lieb-inhom}
   \rho(\la,\xi) + \int\limits_{-\Lambda}^\Lambda K(\la-\mu)\,\rho(\mu,\xi)\,\dd\mu
   = \frac{i}{\pi} t(\la,\xi),
\end{equation}
with
\begin{equation}
   t(\la,\xi)=\frac{\sh\eta}{\sh(\la-\xi)\,\sh(\la-\xi+\eta)}.
\end{equation}
It is easy to see that $\rho(-\la,\eta-\xi)=\rho(\la,\xi)$.
Then the function
\begin{equation}
   \rho_{\mathrm{tot}}(\la)=\frac{1}{2M}\sum_{k=1}^M
             \big[ \rho(\la,\xi_k)+\rho(\la,\eta-\xi_k) \big]
\end{equation}
satisfies the integral equations~\eqref{Lieb} and \eqref{Liebbulk} in the inhomogeneous case, and tends to the ground state
density $\rho$ in the homogeneous limit.
The inhomogeneous integral equation~\eqref{lieb-inhom} can be solved explicitely as
\begin{equation*}
   \rho(\la,\xi)=\begin{cases}
      \displaystyle{\frac{i}{\zeta\,\sh\frac{\pi}{\zeta}(\la-\xi)}}\quad
       &\text{in the massless regime,}\\
      -\displaystyle{\frac{1}{\pi}\prod\limits_{n=1}^\infty \left(\frac{1-q^{2n}}{1+q^{2n}}\right)^2
      \frac{\theta_2(i(\lambda-\xi),q)}{\theta_1(i(\lambda-\xi),q)}},
   \quad q=e^\eta,
       &\text{in the massive regime.}
      \end{cases}
\end{equation*}
Finally we would like to stress that all the functions in \eqref{lieb-inhom} are
holomorphic in a symmetric strip of width $\eta$ around the interval $[-\Lambda,\Lambda]$.
Therefore this equation still holds at the extra root $\check\la$ when it exists.

\section{Scalar products of boundary states}
\label{sec-sc-product}

In order to compute correlation functions following the method
proposed in \cite{KitMT99}, \cite{KitMT00}, it is necessary to have
an explicit expression of the scalar products between a Bethe state
and a general state. In the bulk case, such scalar products have
been represented in the form of a determinant of usual functions in
\cite{Sla89}, \cite{KitMT99}. The method proposed in \cite{KitMT99}
has been used in \cite{Wan02} to obtain a similar representation for
the open XXX chain. In this section, we give the explicit
expressions of the scalar products of boundary states in the $XXZ$
case, and explain briefly how to derive them.

\subsection{Partition functions}
\label{sec-partition}

It is useful, as a starting point for the computation of scalar products, to consider the following functions:
\begin{align}
  &\mathcal{Z}_{M}^{\mathcal{B}_{\pm }}(\{\lambda _{\alpha }\},\{\xi_{k}\};\xi_\pm)
  =\bra{\overline{0}}\,\mathcal{B}_{\pm }(\lambda _{M})\ldots
                          \mathcal{B}_{\pm }(\lambda_{1})\,\ket{0},
 \label{Z_N}\\
 &\mathcal{Z}_{M}^{\mathcal{C}_{\pm }}(\{\lambda _{\alpha }\},\{\xi_{k}\};\xi_\pm)
 =\bra{ 0}\,\mathcal{C}_{\pm }(\lambda _{M})\ldots
                          \mathcal{C}_{\pm}(\lambda_{1})\,\ket{\overline{0}},
\end{align}
where $\ket{\overline{0}}$ is the reference state with all spin down and $\bra{\overline{0}}$ is its dual. Note that
these functions correspond to the partition functions of the six-vertex model with domain wall boundary conditions and
one reflecting end.

\begin{prop}
\label{prop 6}
The above partition functions are related to each others in the following way:
\begin{align}
  &\mathcal{Z}_{M}^{\mathcal{B}_{\pm }}(\{\lambda _{\alpha }\},\{\xi_{k}\};\xi_\pm)
    =( -1)^{M} \mathcal{Z}_{M}^{\mathcal{C}_{\pm}}(\{\lambda _{\alpha }\},\{\xi _{k}\};-\xi_\pm),
      \label{Z-rel-1}
      \\
  &\mathcal{Z}_{M}^{\mathcal{B}_{+}}(\{\lambda _{\alpha }\},\{\xi_{k}\};\xi_+)
    =( -1)^{M} 
    \mathcal{Z}_{M}^{\mathcal{C}_{-}}(\{-\lambda _{\alpha }\},\{\xi _{k}\};\xi_+),
     \label{Z-rel-2}\\
  &\mathcal{Z}_{M}^{\mathcal{C}_{+}}(\{\lambda _{\alpha }\},\{\xi_{k}\};\xi_+)
    =( -1)^{M} 
    \mathcal{Z}_{M}^{\mathcal{B}_{-}}(\{-\lambda _{\alpha }\},\{\xi _{k}\};\xi_+).
      \label{Z-rel-3}
\end{align}
\end{prop}

\textit{Proof --- \ }
As $\Gamma _{x}\,\ket{0} =\ket{\overline{0}}$ and
$\bra{\overline{0}}\,\Gamma _{x}=\bra{0}$,
the relations (\ref{Z-rel-1}) are direct consequences of Lemma \ref{prop 1}.
The other two identities can be proved using the boundary-bulk decomposition of Proposition \ref{prop 3}.
\qed

\begin{prop}
\label{prop 7} \cite{Tsu98}
The partition function $\mathcal{Z}_{M}^{\mathcal{C}_{-}}$ can be represented
as the determinant
\begin{multline}
  \mathcal{Z}_{M}^{\mathcal{C}_{-}}(\{\lambda _{\alpha }\},\{\xi _{k}\};\xi_-)
  =  \pl_{\beta=1}^M \big[\gamma(\la_\beta)\,  a(\la_\beta)\, a(-\la_\beta)\big]\\
  \times\frac{
   \overset{M}{\underset{\beta =1}{\prod }}\overset{M}{\underset{k=1}{\prod }}
         \big[\sinh (\lambda _{\beta }-\xi _{k})\sinh (\lambda _{\beta }+\xi _{k})\big]}
        {\underset{\beta <\gamma }{\prod }\big[
     \sinh \lambda _{\beta \gamma }\sinh \overline{\lambda }_{\beta \gamma }\big]
     \underset{r<s}{\prod }\big[\sinh \xi_{sr}\sinh (\overline{\xi }_{sr}-\eta )\big]}\
   \underset{M}{\det}\mathcal{N}^{\mathcal{C}_{-}}(\la_\alpha,\xi_k;\xi_-),
   \label{Partition-C-}
\end{multline}
where
\begin{equation}
\mathcal{N}_{\alpha k}^{\mathcal{C}_{-}}
  =\frac{\sinh \eta\, \sinh (2\lambda_{\alpha }-\eta )\,\sinh (\xi_{-}+\xi_{k}-\eta /2)}
        {\sinh (\lambda_{\alpha}-\xi_{k}+\eta )\,\sinh (\lambda_{\alpha}+\xi_{k}-\eta)\,
     \sinh (\lambda_{\alpha}-\xi_{k})\,\sinh (\lambda_{\alpha}+\xi_{k})}.
      \label{N^C-}
\end{equation}
\end{prop}

\textit{Proof --- \ }
In \cite{Tsu98}, both a set of recursion relations
and the corresponding solutions were obtained.

We propose in Appendix~\ref{append-part} an alternate derivation of
this representation. It is based, just as in \cite{KitMT99}, on
direct calculations in the basis induced by the twist $F$ introduced
in \cite{MaiS00}. \qed

\subsection{Scalar products}
\label{subsec-sc-product}

Let us define, for two sets of complex variables $\{\la_1,\ldots,\la_N\}$ and $\{\mu_1,\ldots,\mu_N\}$, the following
different scalar products:
\begin{equation}
   \cS_N^{\eps_1,\eps_2} (\{\la\};\{\mu\})
   =\bracket{\psi_{\eps_1}(\{\la\})\, | \, \psi_{\eps_2}(\{\mu\})},
\end{equation}
for $\eps_1,\eps_2\in\{+,-\}$.

\begin{thm}
\label{theo}
Let $\{\lambda _{1},\ldots,\lambda _{N}\}$ be a
solution of the system of Bethe equations\;(\ref{eqBethe}) and
$\{\mu_{1},\ldots,\mu _{N}\}$ be generic complex numbers.
Then, the scalar products between the state $\ket{\psi_+(\{\mu\})}$ and
the eigenstate $\bra{\psi_-(\{\la\})}$, and between the state $\bra{\psi_+(\{\mu\})}$ and
the eigenstate $\ket{\psi_-(\{\la\})}$, are respectively given as
\begin{align}
   &\cS_N^{-,+} (\{\la\};\{\mu\})
   =\prod\limits_{a=1}^{N}\big[\gamma(\la_a)\,d(\la_a)\, d(-\la_a)\big]\
    \frac{\det_N \eT(\{\lambda \},\{\mu \})}{\det_N \eV(\{\lambda \},\{\mu \})},
             \label{C-B+scalar-product}\\
   &\cS_N^{+,-} (\{\mu\};\{\la\})
   =\prod\limits_{a=1}^{N}\big[\gamma(\la_a)\,a(\lambda_a)\, a(-\lambda_a)\big]\
    \frac{\det_N \eT(\{\lambda \},\{\mu \})}{\det_N \eV(\{\lambda \},\{\mu\})},
             \label{C+B-scalar-product}
\end{align}
where the matrices $\eT$ and $\eV$ are defined as
\begin{align}
  &\eT_{\alpha \beta}(\{\lambda \},\{\mu \})
  = \frac{\partial}{\partial\la_\alpha}\tau(\mu_\beta,\{\la\})
               \label{jacobT}\\
  &\eV_{\alpha \beta}(\{\lambda \},\{\mu \})
  = \frac{\sh(2\la_\alpha)\,\sh(2\mu_\beta-\eta)}
         {\sh(2\la_\alpha-\eta)\,
      \sh(\mu_\beta-\la_\alpha)\,\sh(\mu_\beta+\la_\alpha)},
           \label{matV}
\end{align}
in which $\tau(\mu_\beta,\{\la\})$ denotes the eigenvalue \eqref{valpropretau} of the transfer matrix $\cT(\mu)$
on a Bethe eigenstate parametrized by $\{\la_1,\ldots,\la_N\}$.
\end{thm}

\textit{Proof --- \ }
Let us for example consider the scalar product \eqref{C-B+scalar-product}.
This formula can be proved following the same procedure as in \cite{KitMT99}
for the bulk case. As the reference state is invariant
under the action of the operator $F$, we can rewrite this scalar  product
in the $F$-basis and use the explicit representations
$\widetilde{\mathcal{C}}_{-}$ and $\widetilde{\mathcal{B}}_{+}$
given in Lemma~\ref{prop 2} for
the operators $\mathcal{C}_{-}$ and $\mathcal{B}_{+}$ in this basis:
\begin{equation}
  \cS_N^{-,+} (\{\la\};\{\mu\})
  =\bra{0}\prod_{a=1}^{N}\widetilde{\mathcal{C}}_{-}(\lambda _{a})
         \prod_{b=1}^{N}\widetilde{\mathcal{B}}_{+}(\mu _{b}) \ket{0} .
\end{equation}
The idea is then to insert, in front of each operator $\widetilde{\mathcal{B}}_{+}$,
a sum over the complete set of spin states $\ket{i_{1},\ldots,i_{m}}$,
where $\ket{i_{1},\ldots,i_{m}}$ is the state
with $m$ spins down in the sites $i_{1},\ldots,i_{m}$ and $M-m$ spins up in
the other sites.
We are thus led to consider intermediate functions of the form
\begin{multline}
  G^{(m)}(\{\lambda _{k}\},\mu _{1},\ldots,\mu_{m},i_{m+1},\ldots,i_{N})\\
  =\bra{0} \prod_{a=1}^{N}\widetilde{\mathcal{C}}_{-}(\lambda_{a})
     \prod_{b=1}^{m}\widetilde{\mathcal{B}}_{+}(\mu _{b})\,
        \ket{i_{m+1},\ldots,i_{N}}.
     \label{def-G(m)}
\end{multline}
which satisfy the following simple recursion relation:
\begin{align}
  &G^{(m)}(\{\lambda _{k}\},\mu _{1},\ldots,\mu _{m},i_{m+1},\ldots,i_{N})\notag\\
  &\hspace{2.5cm}=\sum\limits_{j\neq i_{m+1},\ldots,i_{N}}
  \bra{j,i_{m+1},\ldots,i_{N}} \,{\widetilde{\cB}}_{+}(\mu _{m})\,
  \ket{i_{m+1},\ldots,i_{N}} \notag\\
  &\hspace{4.5cm}\times
  G^{(m-1)}(\{\lambda _{k}\},\mu_{1},\ldots,\mu_{m-1},j,i_{m+1},\ldots,i_{N}),
   \label{recur1}
\end{align}
Note that the last of this function is precisely the scalar product we want to compute,
\begin{equation}
   G^{(N)}(\{\lambda _{k}\},\mu _{1},\ldots,\mu_{N})
   =\bracket{\psi_-(\{\la\})\, | \, \psi_+(\{\mu\})},
\end{equation}
whereas the first one,
\begin{equation*}
  G^{(0)}(\{\lambda _{k}\},i_{1},\ldots,i_{N})
  =\bra{0} \prod_{a=1}^{N} {\widetilde{\cC}}_{-}(\lambda _{a})\,
  \ket{i_{1},\ldots,i_{N}},
\end{equation*}
is closely related to the partition function
computed in the previous section:
\begin{multline}\label{G0}
 G^{(0)}(\{\lambda _{k}\},i_{1},\ldots,i_{N})
 =\prod_{l\neq i_1,\ldots,i_N} \bigg\{
  \prod_{\alpha=1}^N \big[ b(\la_\alpha-\xi_l)\,
                           b(-\la_\alpha-\xi_l)\big]
  \prod_{\beta=1}^N b^{-1}(\xi_{i_\beta}-\xi_l)\bigg\}\\
  \times \mathcal{Z}_{N}^{\mathcal{C}_{-}}
( \{\lambda _{1},\ldots,\lambda _{N}\},\{\xi _{i_{1}},\ldots,\xi_{i_{N}}\};\xi_-).
\end{multline}

Solving the recursion \eqref{recur1} we obtain, when particularizing the result to the case $m=N$,
\begin{multline}
\bra{0}\prod_{a=1}^{N}\mathcal{C}_{-}(\lambda _{a})
         \prod_{b=1}^{N}\mathcal{B}_{+}(\mu _{b}) \ket{0}
  =\frac{\sh^N\eta \prod\limits_{a=1}^{N}
          \big[\gamma(\la_a)\,d(\lambda_{a})\, d(-\lambda _{a})\big]}
    {\prod\limits_{a>b}^{N}
    \big[\sinh \lambda_{ab}\,\sinh \overline{\lambda }_{ab}\,
        \sinh \mu _{ba}\,\sinh \overline{\mu }_{ba}\big]}\\
  \times \prod_{b=1}^N\frac{\sh(2\la_b-\eta)\,\sh(2\mu_b+\eta)}{\sh(2\mu_b)}
             \ \underset{N}{\det} H_{\alpha \beta}(\{\lambda \},\{\mu \})   ,
\end{multline}
with
\begin{equation}
H_{\alpha \beta }(\{\lambda \},\{\mu \})
  =\frac{ \gamma(\mu_\beta)\big\{ y_{\alpha }(\mu_{\beta };\{\lambda \})
            -y_{\alpha }(-\mu_{\beta};\{\lambda \})\big\} }
    {\sinh (\lambda _{\alpha }-\mu _{\beta })\,
     \sinh (\lambda _{\alpha }+\mu_{\beta })}.
   \label{Matrix-elem-scalarproduct}
\end{equation}
This ends the proof of (\ref{C-B+scalar-product}).

As for (\ref{C+B-scalar-product}), it can be proved
following the same procedure, provided one writes the operators $\mathcal{C}_{+}$ and $\mathcal{B}_{-}$ in
the $\overline{F}$-basis as in Lemma \ref{prop 2}.
\qed

All the remaining scalar products can be expressed in terms
of those given in Theorem \ref{theo}. Indeed, we have the following corollary.

\begin{cor}\label{cor}
Let $\{\lambda _{1},\ldots,\lambda _{N}\}$ be a solution of
the system of Bethe equations (\ref{eqBethe}) and $\{\mu _{1},\ldots,\mu _{N}\}$ be generic complex numbers. Then,
\begin{align}
  &\cS_N^{-,+} (\{\mu\};\{\la\})=  \cS_N^{-,+} (\{-\la\};\{-\mu\}),
       \label{C-B+scalar-product-2}\\
  &\cS_N^{+,-} (\{\la\};\{\mu\}) =\cS_N^{+,-} (\{-\mu\};\{-\la\}).
       \label{C+B-scalar-product-2}
\end{align}
Finally, the proportionality between $\pm $ Bethe
eigenstates, given in Proposition \ref{prop 5}, allows us to complete the
list of scalar products where one of the boundary states is an eigenstate.
\end{cor}

\textit{Proof --- \ }
The idea is to go to the $F$-basis, and then to insert, between the product of the operators $\widetilde{\mathcal{C}}$
and $\widetilde{\mathcal{B}}$, the identity as a sum over convenient intermediate spin states,
 and finally to use
the results of Proposition \ref{prop 6}.

For example, applying this procedure to the left hand side of (\ref{C-B+scalar-product-2}) in the $F$-basis, one obtains
the relation
\begin{multline}
  \bra{0}\prod_{b=1}^{N}\mathcal{C}_{-}(\mu _{b})
  \prod_{a=1}^{N}\mathcal{B}_{+}(\lambda _{a})\,\ket{0}
   =\bra{0}\prod_{a=1}^{N}\mathcal{C}_{-}(-\lambda _{a};\xi_+)
   \prod_{b=1}^{N}\mathcal{B}_{+}(-\mu _{b};\xi_-)\,\ket{0},
   \label{rel-corollary-0}
\end{multline}
which holds for any arbitrary sets of complex numbers
$\{\lambda _{a}\}$ and $\{\mu _{b}\}$. The
Bethe equations and the scalar product formula (\ref
{C-B+scalar-product}) being invariant under the simultaneous exchanges $\xi
_{+}\rightarrow \xi _{-}$ and $\xi _{-}\rightarrow \xi _{+}$, we obtain,
under the additionnal assumption that $\{\lambda _{a}\}$ is a solution of the system of Bethe equations
and thanks to Proposition \ref{prop 4}, that the scalar
product on the right hand side of (\ref{rel-corollary-0}) is given by (\ref
{C-B+scalar-product}) evaluated at $\{-\lambda _{a}\}$ and $\{-\mu _{b}\}$.
\qed

\subsection{The Gaudin formula and the orthogonality of Bethe states}
\label{sec-gaudin}

The scalar product formul\ae\ derived in the previous section can be used,
as in the bulk case, to compute the norm of boundary Bethe states and
prove the orthogonality of eigenstates corresponding to different solutions
of the Bethe equations.

\begin{cor}\label{cor-ortho}
Let $\{\lambda _{1},\ldots,\lambda _{N}\}$ and $\{\mu_{1},\ldots,\mu_{N}\}$
be two different solutions of the system of the Bethe equations, that is,
\begin{equation}
\{\sigma _{1}\lambda _{1},\ldots,\sigma _{N}\lambda _{N}\}\neq
\{\mu_{1},\ldots,\mu_{N}\}\quad \text{for each }\ \sigma _{j}=\pm,\ j=1,\ldots,N.
\end{equation}
Then, for $\eps_1,\eps_2\in\{+,-\}$, the scalar product $\cS_N^{\eps_1,\eps_2} (\{\la\};\{\mu\})$ vanishes, i.e.
the corresponding Bethe states $\bra{\psi_{\eps_1}(\{\mu\})}$ and $\ket{\psi_{\eps_2}(\{\la\})}$ are orthogonal.
\end{cor}

\textit{Proof --- \ }
Let us show the orthogonality of the Bethe states corresponding to two
different solutions $\{\lambda _{1},\ldots,\lambda _{N}\}$\ and
$\{\mu_{1},\ldots,\mu_{N}\}$ of \eqref{eqBethe}. In such a case, the scalar products $\cS_N^{\eps_1,\eps_2} (\{\la\};\{\mu\})$
are proportional to each others for the different values of $\eps_1,\eps_2$, and the orthogonality follows from the fact that the
determinant in (\ref{C-B+scalar-product})\ is equal to zero. Indeed,
there exists a non-trivial vector $v(\{\la\},\{\mu\})$ such that:
\begin{equation}
\sum\limits_{k=1}^{N} H_{jk} (\{\la\},\{\mu\})\, v_{k}(\{\la\},\{\mu\})=0,
            \text{ for any }j=1,\ldots,N.
\label{linear-dependence}
\end{equation}
Such a vector $v(\{\la\},\{\mu\})$ can be constructed in the following way:
\begin{equation}\label{vecteur-v}
v_{j}(\{\la\},\{\mu\})=\frac{\prod\limits_{k=1}^{N}\sinh (\lambda_{j}-\mu_{k})
            \prod\limits_{k=1}^{N}\sinh (\lambda _{j}+\mu _{k})}
       {\prod\limits_{k\neq j}\sinh \lambda _{jk}
        \prod\limits_{k\neq j}\sinh \overline{\lambda}_{jk}}.
\end{equation}
To check that the equations (\ref{linear-dependence})\ are satisfied with the vector \eqref{vecteur-v}, we
use the explicit expression for the matrix elements
$H_{jk}(\{\la\},\{\mu\})$ $(\ref{Matrix-elem-scalarproduct})$ and apply the Bethe equations
for the solution $\{\mu_{1},\ldots,\mu_{N}\}$.
\qed

\begin{cor}
Let $\{\la_1,\ldots,\la_N\}$ be a solution of the system of Bethe equations.
We have
\begin{multline}
 \cS_N^{-,+} (\{\la\};\{\la\})
    = \sinh^N \eta\frac{\pl_{a=1}^N \big[\gamma^2(\la_a)\,d(\la_a)\,d(-\la_a)\,
             y_{a}(-\lambda_{a};\{\lambda \})\big]}
          {\prod\limits_{a\neq b}^{N}\big[\sinh \lambda _{ab}\,
       \sinh \overline{\lambda }_{ab}\big]} \\
  \times
      \prod_{j=1}^N\frac{\sinh(2\lambda_{j}-\eta )\,\sinh (2\lambda_{j}+\eta )}
                        {\sinh^{2}(2\lambda_{j})}\,
      \underset{N}{\det} \Phi_{jk}'(\{\la\}).
      \label{Gaudin}
\end{multline}
Here $\Phi'$ is the Gaudin matrix:
\begin{equation}
  \Phi'_{jk}(\{\la\})=\frac{\partial }{\partial \lambda _{j}}
    \log \frac{y_{k}(-\lambda _{k};\{\lambda \})}{y_{k}(\lambda _{k};\{\lambda \})},
\end{equation}
with $y_{j}(x;\{\lambda \})$ defined as in (\ref{fun_y}).
The norm of the corresponding Bethe eigenstate follows then from Proposition~\ref{prop 5} and Proposition~\ref{Dual}.
\end{cor}

\subsection{Partial scalar products in the thermodynamic limit}\label{sec-part-sc}

For the computation of correlation functions, it is usefull to have an expression for
partial renormalized scalar products. For some partition $\alpha_+\cup\alpha_-$ of
$\intn{1}{N}$, we consider the sets of variables $\{\la_1,\ldots,\la_N\}$ and
$\{\mu_1,\ldots,\mu_N\}$, with $\{\la\}$ solution of the system of Bethe equations \eqref{eqBethe}, such that
\beq
 \paa{\la}=\paa{\la_a}_{a\in\alpha_-}\cup\paa{\la_b}_{b\in\alpha_+},\qquad
 \paa{\mu}=\paa{\la_a}_{a\in\alpha_-}\cup\paa{\xi_{i_b}}_{b\in\alpha_+},
\enq
in which $\{\xi_{i_b}\}_{b\in\alpha_+}$ is a subset of $\{\xi_1,\ldots,\xi_M\}$.
Then,
\begin{multline*}
 \f{\cS_N^{+,+} (\{\la\};\{\mu\})}{\cS_N^{+,+} (\{\la\};\{\la\})}=
 \pl_{b\in\alpha_+}
 \frac{\gamma(\la_b)\,\sh(2\la_b)\,\sh(2\la_b-\eta)\,\sh(2\xi_{i_b}+\eta)\,
                 \hat{y}(\xi_{i_b};\{\la\})}
      {\gamma(\xi_{i_b})\,\sh(2\xi_{i_b})\,\sh(2\xi_{i_b}-\eta)\,\sh(2\la_b+\eta)\,
                 \hat{y}(\la_b;\{\la\})}
      \\
 \times\prod_{\substack{a,b\in\alpha_+ \\ a>b}}
  \frac{\sh(\la_{ba})\,\sh(\bar\la_{ba})}
       {\sh(\xi_{i_b\, i_a})\,\sh(\bar\xi_{i_b\, i_a})}
 \prod_{\substack{a\in\alpha_-\\ b\in\alpha_+}}
  \frac{\sh(\la_{ba})\,\sh(\bar\la_{ba})}
       {\sh(\xi_{i_b}-\la_a)\, \sh(\xi_{i_b}+\la_a)}
  \frac{\det_N \mathcal{M}}{\det_N \mathcal{N}},
\end{multline*}
with $\hat{y}$ given by \eqref{fun_y} and
\begin{align}
  &\mathcal{N}_{a b}= 2M\delta_{a b} \Big\{ p´(\la_a)
  +\frac{1}{2M} g´(\la_a;\xi_+,\xi_-) - \frac{\pi}{M}
        \sum_{k=1}^N\big[ K(\la_a -\la_k) + K(\la_a +\la_k)\big]
              \nonumber\\
  &\hspace{4.2cm}  +\frac{2\pi}{M} K(2\la_a) \Big\}
     + 2\pi \big[ K(\la_a-\la_b)- K(\la_a + \la_b)\big],\\
  &\mathcal{M}_{a b}=
     \begin{cases}
           \mathcal{N}_{ab} & \text{if }b\in\alpha_-,\\
           i\big[ t(\la_a,\xi_{i_b})-t(\la_a,\eta-\xi_{i_b}) \big]
                            & \text{if }b\in\alpha_+.
     \end{cases}
\end{align}
It remains to caracterize the ration of the two determinants of $\mathcal{M}$ and
$\mathcal{N}$, wich reduces to the determinant of a matrix $\mathcal{S}$ of size
$|\alpha_+|$:
\begin{equation}
   \frac{\det_N\mathcal{M}}{\det_N\mathcal{N}}
         =\det_{a,b\in\alpha_+}\mathcal{S}_{ab}, \qquad
   \text{with}\quad \mathcal{S}_{ab}
     =\sum_{\beta=1}^N\big(\mathcal{N}^{-1}\big)_{a\beta}\,\mathcal{M}_{\beta b}.
\end{equation}
It is actually possible, as in the bulk case, to compute explicitely $\mathcal{S}_{ab}$ for the ground state
in the thermodynamic limit.

Indeed, it is easy to see that, if  $\la_j$ corresponds to a real root $\alpha_j$,
\begin{equation}\label{N-reel}
  \sul_{\substack{p=1 \\ \alpha_p\ \text{real}}}^{N} \mathcal{N}_{j p}
   \frac{\rho(\la_p,\xi_k)-\rho(\la_p,\eta-\xi_k)}{2M\rho(\la_p)}
  \underset{M\tend\infty}{\longrightarrow}
   i\big[ t(\la_j,\xi_k)-t(\la_j,\eta-\xi_k) \big].
\end{equation}
This follows from the fact that, if $\la_j$ corresponds to a real root of the ground state, the matrix element
$\mathcal{N}_{j p}$ can be expressed as
\begin{equation}
   \mathcal{N}_{j p} = 2M\delta_{j p} \Big\{ \rho(\la_j)
         + O\Big(\frac{1}{M}\Big)\Big\}
        +2\pi\big[ K(\la_j-\la_p)-K(\la_j+\la_p)\big],
\end{equation}
from the symmetry property $\rho(\la,\mu)=\rho(-\la,\eta-\mu)$ of the inhomogeneous density, and from the inhomogeneous
integral equation \eqref{lieb-inhom}.

Therefore, if the ground state contains only real roots (i.e. in the case $\tilde\xi_-<0$ or $\tilde\xi_->\zeta/2$),
\begin{equation}
  \mathcal{S}_{ab}\underset{M\tend\infty}{\ttend}
  \frac{\rho(\la,\xi_{i_b})-\rho(\la_a,\eta-\xi_{i_b})}{2M\rho(\la_a)}.
\end{equation}

Let us now consider the ground state in the case $0<\tilde\xi_-<\zeta/2$, and let $\la_1=\check\la$ corresponding to
the complex root (i.e. $\la_1\underset{M\tend\infty}{\ttend} \eta/2-\xi_-$); then, there exists $\gamma>0$ such that
\begin{equation}\label{N1p}
   \mathcal{N}_{1 p} = g'(\check\la) \Big\{\delta_{1 p}
         \big[ 1 + O\big(M e^{-\gamma M}\big)\big] + O\big( e^{-\gamma M} \big) \Big\},
\end{equation}
with $[g'(\check\la)]^{-1}\sim i\sh(\check\la-\xi_-+\eta/2)$ in the large $M$ limit. Therefore, using \eqref{N-reel}, the
inhomogeneous integral equation \eqref{lieb-inhom} at the point $\check\la$ and the estimation of $(\mathcal{N}^{-1})_{a 1}$
following from \eqref{N1p}, we obtain that
\begin{align}
   \mathcal{S}_{ab}
   &=\big(\mathcal{N}^{-1}\big){}_{a1}\, \mathcal{M}_{1b}
  +\sum_{\beta=2}^N \big(\mathcal{N}^{-1}\big){}_{a\beta}\, \mathcal{M}_{\beta b},
                     \\ \nonumber
   &\underset{M\tend\infty}{\sim}
       \begin{cases}
          \displaystyle{i\pi \sh\big(\check\la-\xi_-+\eta/2\big)
            \big[ \rho(\check\la,\xi_{i_b})-\rho(\check\la,\eta-\xi_{i_b})\big] }
            &\text{ if $a=1$,}\smallskip \\
          \displaystyle{\frac{ \rho(\la_a,\xi_{i_b})-\rho(\la_a,\eta-\xi_{i_b})}{
          2M\rho(\la_a)}}            &\text{ if $a\ne 1$.}
       \end{cases}
\end{align}

\section{Action of local operators on a boundary state}
\label{sec-act-state}

In order to compute correlation function, one should now determine the action of the corresponding
local operators on a boundary state. As in the bulk case \cite{KitMT99}, the idea is to solve the inverse
scattering problem, i.e. to express local operators in terms of the generators of the Yang-Baxter algebra.

A natural idea would be to try to express these local operators directly in terms of the generators
$\cA_+$, $\cB_+$, $\cC_+$, $\cD_+$ (or $\cA_-$, $\cB_-$, $\cC_-$, $\cD_-$) of the {\em boundary}
Yang-Baxter algebra. However, although it is quite easy to reconstruct in such a way a local spin
operator at the first site of the chain \cite{Wan00}, it seems much more difficult, due to the lack
of translation invariance, to obtain effective formulas on the other sites of the chain. In pratice, the
reconstruction proposed in \cite{Wan00}, which involves the adjoint action of the bulk translation
operators $(A+D)(\xi_k)$, is unadapted to compute correlation functions of the boundary model since
eigenstates of the Hamiltonian are no more eigenstates of these translation operators.

Quite surprisingly, it is actually possible to use directly a version of the {\em bulk} inverse problem
to compute the action of local operators on a {\em boundary} state. In this section, we will show how to
reformulate the bulk inverse problem so as to circumvent the use of the translational invariance of the chain.
Then, using Proposition~\ref{prop 3}, we will act with the corresponding products of bulk operators
on a boundary state, obtaining a sum over some bulk states that eventually reduces to a sum over boundary states.

\subsection{The bulk inverse problem revisited}
\label{sec-inv}

Let us define a familly of algebra homomorphisms $\chi_i:\eH\tend\eH$ as
\beq
  \chi_i: X \mapsto \widehat{R}_{i\, i-1} \dots \widehat{R}_{i\, 1} X
           \widehat{R}_{1 \, i} \dots \widehat{R}_{i-1 \, i}.\label{defchi}
\enq
Then, for a local operator $X_i$ at site $i$ (i.e. which acts non trivially only on $\eH_i$), $\chi_i (X_i)$ can be
expressed in terms of the bulk monodromy matrix entries as
\begin{align}
\chi_i(X_i)&=\big[a(\xi_i)\, d(\xi_i-\eta)\big]^{-1}\,
    \text{tr}_{0}\big[T_0 (\xi_i)\, X_0\big]\,(A+D)(\xi_i-\eta),
                 \label{chi_i1}\\
   &=\big[a(\xi_i)\, d(\xi_i-\eta)\big]^{-1}\, (A+D)(\xi_i)\,
       \text{tr}_{0}\big[\sg_0^{y}T_0^{t_0}(\xi_i-\eta)\sg^{y}_0\, X_0\big].
\label{chi_i2}
\end{align}
To compute the bulk correlation functions, the authors of \cite{KitMT99}, \cite{KitMT00} used the
reconstruction \eqref{chi_i1}\footnote{Note that we express here the result in a slightly different
form, using that $\big[(A+D)(\xi_i)\big]^{-1}=\big[a(\xi_i)\, d(\xi_i-\eta)\big]^{-1}\, (A+D)(\xi_i-\eta)$.}, together
with the fact that
\begin{equation}\label{reconstr-bulk}
     X_i=\prod_{\alpha=1}^{i-1}(A+D)(\xi_\alpha)\, \chi_i(X_i)\,
         \prod_{\alpha=1}^{i-1}\big[(A+D)(\xi_\alpha)\big]^{-1}.
\end{equation}
This was convenient there because the product of bulk transfer matrices merely produces a numerical factor
when applied on a bulk Bethe state. As it is no longer the case when applied on a boundary Bethe state, the
strategy is to simplify this product instead.

We can make the following observation:

\begin{lemma}\label{lemme-prod-op}
   The products of the bulk monodromy matrix elements $T_{\eps\,\eps'}(\xi_i)\,
T_{\bar{\eps}\, \bar{\eps}'}(\xi_i-\eta)$ and $T_{\eps'\,\eps}(\xi_i-\eta)\,
T_{\bar{\eps}'\, \bar{\eps}}(\xi_i)$ vanish if $\eps=\bar\eps$.
\end{lemma}

\Proof It follows directly from the fact that
$
 \chi_i\big(E^{\mu'\,\mu}_{i}\big)\,\chi_i\big(E^{\bar\mu\,\bar\mu'}_i\big)
        =\delta_{\mu,\bar\mu} \, \chi_i\big(E^{\mu'\, \bar\mu'}_i\big)
$
and from the expressions \eqref{chi_i1}, \eqref{chi_i2} of
$\chi_i\big(E^{\mu'\,\mu}_{i}\big)$, $\chi_i\big(E^{\bar\mu\,\bar\mu'}_{i}\big)$ in terms of the bulk monodromy matrix,
in which $E^{\mu\,\mu'}_{i}$ denotes the elementary matrix at site $i$ with elements
$\big( E^{\mu\,\mu'}_{i}\big)_{a\, b}=\delta_{a,\mu}\, \delta_{b,\mu'}$.
\qed

\begin{rem}\label{rem-identity}
Other interesting identities may be proved in the same way. For example, from the fact that
$\chi_i\big(E^{12}_{i}\big)\,\chi_i\big(E^{21}_i\big)= \chi_i\big(E^{11}_{i}\big)\,\chi_i\big(E^{11}_i\big)$, one
obtains that $C(\xi_i)B(\xi_i-\eta)=-A(\xi_i)D(\xi_i-\eta)$.
\end{rem}

This result can be generalized to a product of $2m$ operator entries of the bulk monodromy matrix:
\begin{thm}
\label{theoremannulationoperteurs}
For any set of inhomogeneity parameters $\{\xi_{i_1},\ldots,\xi_{i_n}\}$, the following product of bulk operators
\beq\label{prod-n}
  T_{\eps_{i_n}\, \eps'_{i_n}}(\xi_{i_n})\dots T_{\eps_{i_1}\, \eps'_{i_1}}(\xi_{i_1})\,
  T_{\bar\eps_{i_1}\, \bar\eps'_{i_1}}(\xi_{i_1}-\eta)\dots
       T_{\bar\eps_{i_n}\,\bar\eps'_{i_n}}(\xi_{i_n}-\eta)
\enq
vanishes if, for some  $k\in\{i_1,\ldots,i_n\}$, $\eps_k=\bar\eps_k$.
\end{thm}

\Proof
It can be proved by recursion on $n$, the case $n=1$ following from Lemma~\ref{lemme-prod-op}.

Let us suppose that the result holds for $n-1$, and consider the product \eqref{prod-n} with $\eps_{i_n}=\bar\eps_{i_n}$.
Using the commutation relation given by the quadratic algebra \eqref{RTT}, one can move the exterior operators
(at position $n$) $T_{\eps_{i_n}\, \eps'_{i_n}}(\xi_{i_n})$, $T_{\bar\eps_{i_n}\,\bar\eps'_{i_n}}(\xi_{i_n}-\eta)$
through those at position $n-1$ (evaluated respectively  at $\xi_{i_{n-1}}$
and $\xi_{i_{n-1}}-\eta$). Considering all possible cases for
$T_{\eps_{i_n}\, \eps'_{i_n}}(\xi_{i_n})$, $T_{\bar\eps_{i_n}\,\bar\eps'_{i_n}}(\xi_{i_n}-\eta)$ and
$T_{\eps_{i_{n-1}}\, \eps'_{i_{n-1}}}(\xi_{i_{n-1}})$, $T_{\bar\eps_{i_{n-1}}\,\bar\eps'_{i_{n-1}}}(\xi_{i_{n-1}}-\eta)$, it
is easy to see that the resulting product of the $2(n-1)$ inner operators should vanish due to the recursion hypothesis.
\qed

This leads to the following corollary concerning the reconstruction of a product of local operators acting on
successive sites of the chain:

\begin{cor}\label{cor-elem}
A product of elementary matrices acting on the first $m$ sites of the chain can be expressed as the following
product of entries of the bulk monodromy matrix:
\begin{multline}
  E_{1}^{\eps_1\,\eps'_1}\dots E_{m}^{\eps_m\,\eps'_m}=
  \pl_{i=1}^{m} \big[a(\xi_i)\, d(\xi_i-\eta)\big]^{-1}
               \\
  \times
  T_{\eps'_1\,\eps_1}(\xi_1)\dots T_{\eps'_m\,\eps_m}(\xi_m)\,
  T_{\bar\eps_m\,\bar\eps_m}(\xi_m-\eta) \dots
  T_{\bar\eps_1\,\bar\eps_1}(\xi_1-\eta)
\label{blockselementaires}
\end{multline}
with $\bar\eps_i=\eps'_i+ 1 \pmod 2$.
\end{cor}

\Proof This is a direct consequence of the solution \eqref{chi_i1}, \eqref{reconstr-bulk} of the inverse problem, of
the fact that $\big[(A+D)(\xi_i)\big]^{-1}=\big[a(\xi_i)\, d(\xi_i-\eta)\big]^{-1}\, (A+D)(\xi_i-\eta)$, and of the previous theorem.
\qed

\subsection{Action on a bulk state}
\label{sec-act-bulk}

Let us now establish the action of a product of elementary matrices of the form \eqref{blockselementaires}
on an arbitrary bulk state $\ket{\{\la_j\}_{1\le j\le N}}=\prod_{j=1}^N B(\la_j)\ket{0}$. We refer for example
to~\cite{KitMT99} for the explicit expression, in our notations, of the action on such a state of a single operator
entry  of the monodromy matrix\footnote{There the action on the left was considered, but the coefficients are the same
when one considers an action on the right instead.}. Let us just recall that, like in \eqref{FCactD}, the action of
$A(\mu)$ or $D(\mu)$ produces two kinds of terms: a {\em direct term}, which leaves the state
untouched, and {\em indirect terms}, resulting in new states with one $\la_j$ replaced by $\mu$.
With this terminology, the action of operators of the form \eqref{blockselementaires} can be computed
using the following lemma:

\begin{lemma}\label{lem-action}
The action on a bulk state $\ket{\{\la_j\}_{1\le j\le N}}$ of a string of operators
\begin{equation}
  \mathcal{O}_{\eps_{i_1},\ldots,\eps_{i_n}}^{\eps'_{i_1},\ldots,\eps'_{i_n}}=
  \underbrace{T_{\eps'_{i_n}\,\eps_{i_n}}(\xi_{i_n})\dots
              T_{\eps'_{i_1}\,\eps_{i_1}}(\xi_{i_1})}_{\pa{1}}
  \underbrace{T_{\bar{\eps}_{i_1}\,\bar{\eps}_{i_1}}(\xi_{i_1}-\eta) \dots
              T_{\bar{\eps}_{i_n}\,\bar{\eps}_{i_n}}(\xi_{i_n}-\eta)}_{\pa{2}}
\end{equation}
with $\bar\eps_l=\eps'_l+1\pmod 2$, has the following properties.
\begin{itemize}
  \item The only non-zero contributions of the tails
 operators $\pa{2}$ come from
\begin{itemize}
 \item[(i)]  the indirect action  of all $A(\xi_l-\eta)$ operators;

 \item[(ii)]  the direct action of all $D(\xi_l-\eta)$ operators.
\end{itemize}
\item In what concerns the head operators $\pa{1}$,
\begin{itemize}
  \item[(iii)]  if $\eps'_l=1$, the action of the operator
$T_{\eps'_l\,\eps_l}(\xi_l)$ (i.e. $A(\xi_l)$ or $B(\xi_l)$) does not result in any substitution of a parameter $\xi_i-\eta$;
 \item[(iv)]  if $\eps'_l=2$,  the action of the operator
$T_{\eps'_l\,\eps_l}(\xi_l)$ (i.e. $D(\xi_l)$ or $C(\xi_l)$) substitutes $\xi_l-\eta$ with $\xi_l$; moreover,
if there were others parameters $\xi_j-\eta$, $j\ne l$, in the initial state, they are still present in the resulting state.
\end{itemize}
\end{itemize}
\end{lemma}

\Proof
{\it (i)} and {\it (iii)} follow from the fact that $a(\xi_l-\eta)=0$.

In order to prove {\it (ii)}, let us consider the action of some operator $D(\xi_{i_l}-\eta)$: its indirect contribution
produces a state of the type
$
    B(\xi_{i_l}-\eta)\ket{\{\mu_j\}_{1\le j\le N-1}},
$
for a certain set of parameters $\{\mu_j\}$. However, Theorem~\ref{theoremannulationoperteurs} guarantees that the
operator product
\begin{equation*}
T_{\eps'_{i_l}\,\eps_{i_l}}(\xi_{i_l})\,
\mathcal{O}_{\eps_{i_1},\ldots,\eps_{i_{l-1}}}^{\eps'_{i_1},\ldots,\eps'_{i_{l-1}}}
B(\xi_{i_l}-\eta)
\end{equation*}
is zero, hence {\it (ii)}.

Let us now prove  {\it (iv)} by induction.
If $\eps'_{i_1}=2$, then $T_{\eps'_{i_1}\,\eps_{i_1}}(\xi_{i_1})$ acts on a state of the form
$B(\xi_{i_1}-\eta)\ket{\{\mu_j\}}$ for a certain set of parameters $\{\mu_j\}$, and is either equal to
\begin{itemize}
\item $D(\xi_{i_1})$, which acts only indirectly, and which can only  replace
$\xi_{i_1}-\eta$ with $\xi_{i_1}$; indeed, any other replacement would produce a state of the form
$B(\xi_{i_1})B(\xi_{i_1}-\eta)\ket{\{\bar{\mu}_j\}}$ (where $\{\bar{\mu}_j\}$ is a subset of $\{\mu_j\}$), which is
zero according to Lemma~\ref{lemme-prod-op}.
\item $C\pa{\xi_i}$, which gives, using Remark~\ref{rem-identity},
\begin{equation}
  C(\xi_{i_1})\,B(\xi_{i_1}-\eta)\ket{\{\mu_j\}}=
  -A(\xi_{i_1})\,D(\xi_{i_1}-\eta)\ket{\{\mu_j\}};
\end{equation}
$D(\xi_{i_1}-\eta)$ acts only directly since $A(\xi_{i_1})B(\xi_{i_1}-\eta)=0$ from Lemma~\ref{lemme-prod-op}, and
$A(\xi_{i_1})$ cannot replace any other $\xi_j-\eta$ since $a(\xi_j-\eta)=0$.
\end{itemize}
If {\it (iv)} is proved for all operators until position $l-1$,
then, from   {\it (i)}, {\it (ii)}, {\it (iii)} and the induction
hypothesis, $T_{\eps'_{i_l}\,\eps_{i_l}}(\xi_{i_l})$ acts on a state
of the form $B(\xi_{i_l}-\eta)\ket{\{\mu_j\}}$ for a certain set of
parameters $\{\mu_j\}$, and the same reasoning as for $l=1$ applies.
\qed

In order to express the action of a product of elementary operators
$\prod_{j=1}^{m}E_{j}^{\eps_{j},\eps_{j}^{\prime }}$ on a bulk state, let us consider the following set of indices:
\begin{alignat*}{2}
  &\beta_{+}=\{j:\,1\leq j\leq m,\,\eps_{j}=1\},\quad &
  &\mathrm{card}(\beta_{+})=s^{\prime },\\
  &\beta_{-}=\{j:\,1\leq j\leq m,\,\eps^{\prime}_{j}=2\},\quad &
  &\mathrm{card}(\beta_{-})=s.
\end{alignat*}
Since our final goal is to compute correlation functions, one considers here only products such that
the total number of indices in the sets $\beta_{+}$ and $\beta_{-}$ is $s+s^{\prime }=m$, as otherwise
the corresponding ground state average value is zero.
We can thus introduce a set of indices $i_{p}\in \{1,\ldots,m\}$ such that
\begin{alignat}{3}
   &\beta_{-}=\{i_{p}\}_{p\in \{1,\ldots,s\}},
        & &\text{ with } i_{k}<i_{h} & &\text{ for } 0<k<h\leq s,\\
   &\beta_{+}=\{i_{p}\}_{p\in \{s+1,\ldots,m\}},
        & &\text{ with } i_{k}>i_{h} & &\text{ for } s<k<h\leq m.
\end{alignat}
From Corollary~\ref{cor-elem} and Lemma~\ref{lem-action}, we obtain the following result:

\begin{prop}\label{prop-act-bulk}
The action of a product of elementary operators on an arbitrary bulk state can be
written as
\begin{equation}
  \prod\limits_{j=1}^{m} E_{j}^{\eps_{j},\eps_{j}^{\prime }}
  \prod\limits_{k=1}^{N} B(\lambda _{k})|\,0\,\rangle
  =\sum\limits_{\beta_{m}}
   \mathcal{F}_{\beta _{m}}(\lambda _{1},\dots ,\lambda _{N+m})
   \prod\limits_{\substack{ k=1 \\ k\notin \beta _{m}}}^{N+m}
          B(\lambda_{k})|\,0\,\rangle ,
\label{prop.1}
\end{equation}
in which we have defined
$
   \lambda _{N+j}:=\xi _{m+1-j}\text{ for }j\in \{1,\ldots,m\}.
$
In (\ref{prop.1}), the sums are over all the sets of $m$ indices $\beta_{m}=\{b_{1},\ldots,b_{m}\}$, where
the $b_{p}$ are defined by
\begin{equation}\label{sum-beta}
\begin{cases}
b_{p}\in \{1,\ldots,N\}\setminus\{b_{1},\ldots,b_{p-1}\}\qquad
         & \text{for }0<p\leq s,  \\
b_{p}\in \{1,\ldots,N+m+1-i_{p}\}\setminus\{b_{1},\ldots,b_{p-1}\}\
         & \text{for }s<p\leq m,
\end{cases}
\end{equation}
and the coefficient $\mathcal{F}_{\beta _{m}}$ is
\begin{align}
  \mathcal{F}_{\beta _{m}}(\{\lambda \})
  =&\prod_{j=1}^{m}\Bigg\{ \frac{a(\lambda _{b_j})}{a(\xi _j)} \
   \frac{\prod\limits_{k=1}^{N}\sinh (\lambda _{k\,b_j}+\eta )}
    {\prod\limits_{\substack{ k=1 \\ k\neq b_j}}^{N}\sinh(\lambda_{k\,b_j})}\
    \prod_{k=1}^{N}\frac{\sinh (\lambda _{k}-\xi _j)}
                        {\sinh (\lambda _{k}-\xi _j+\eta )}\Bigg\}
           \nonumber\\
    &\qquad\times
   \prod_{1\leq i<j\leq m}\frac{\sinh (\lambda _{b_{i}\,b_{j}})}
                               {\sinh (\lambda _{b_{i}\,b_{j}}+\eta )}\
   \prod_{p=1}^{s}
      \frac{\prod\limits_{k=i_{p}+1}^{m}\sinh (\lambda _{b_{p}}-\xi _{k}+\eta )}
           {\prod\limits_{k=i_{p}}^{m}\sinh (\lambda_{b_{p}}-\xi _{k})}
           \nonumber\\
    &\qquad\times \prod_{p=s+1}^{m}
      \frac{\prod\limits_{k=i_{p}+1}^{m}\sinh (\xi _{k}-\lambda _{b_{p}}+\eta )}
           {\prod\limits_{\substack{ k=i_{p} \\ k\neq N+m+1-b_{p}}}^{m}
            \sinh (\xi _{k}-\lambda_{b_{p}})} .
\end{align}
\end{prop}

Let us point out that, if the parameters $\la_1,\ldots,\la_N$ are solutions of the bulk
Bethe equations, such a result agrees with what can be obtained with the method used in
\cite{KitMT00}\footnote{taking into account that we consider here an action to the right,
whereas in \cite{KitMT00} we considered an action to the left.}.

\subsection{Action on a boundary state}
\label{sec-act-boundary}

We use the decomposition \eqref{B_b-vector},\eqref{HB+} of boundary states into bulk ones in order
to compute the action of a string of elementary operators on an
arbitrary boundary state constructed from $\mathcal{B}_+$ operators.
It is remarkable that we are eventually able to express explicitely the result as a linear
combination of such boundary states.
Indeed, using the same notations as in Proposition~\ref{prop-act-bulk}, we have,

\begin{prop}\label{prop-act-bound}
The action of a product of elementary operators on a boundary state can be
written as:
\begin{equation}
  \prod\limits_{j=1}^{m}E_{j}^{\eps _{j},\eps_{j}^{\prime }}
   \prod\limits_{k=1}^{N}\mathcal{B}_{+}(\lambda _{k})|\,0\,\rangle
  =\sum\limits_{\beta _{m}}\mathcal{F}_{\beta _{m}}^{+}(\{\lambda\})
   \prod\limits_{\substack{ k=1  \\ k\notin \beta _{m}}}^{N+m}
              \mathcal{B}_{+}(\lambda _{k})|\,0\,\rangle ,
\end{equation}
with $\beta_m$ defined as in \eqref{sum-beta} and the coefficient $\mathcal{F}_{\beta _{m}}^{+}$ given as
\begin{align}
   \mathcal{F}_{\beta _{m}}^{+}(\{\lambda \})
   &=\sum\limits_{\sigma_{\alpha_+}=\pm}
   \prod\limits_{j=1}^{m}\frac{a(\lambda _{b_{j}}^{\sigma })}{a(\xi _{j})}\
   \frac{H_{\sigma _{\alpha _{+}}}^{\mathcal{B}_{+}}(\{\lambda_{\alpha _{+}}\})}
        {H_{1}^{\mathcal{B}_{+}}(\{\xi _{\gamma _{+}}\})}
   \prod\limits_{1\leq i<j\leq m}
   \frac{\sinh\lambda_{b_{i}b_{j}}^{\sigma}}{\sinh(\lambda_{b_{i}b_{j}}^{\sigma}+\eta)}
        \notag \\
        &\times
  \prod\limits_{i\in \alpha _{-}}\prod\limits_{\epsilon=\pm}\bigg\{
  \prod\limits_{j\in \alpha _{+}}
  \frac{\sinh (\lambda _{j}^{\sigma }+\epsilon \lambda _{i}-\eta )}
       {\sinh (\lambda _{j}^{\sigma }+\epsilon \lambda _{i})}
  \prod\limits_{j\in \gamma _{+}}
  \frac{\sinh (\xi _{j}+\eps \lambda _{i})}{\sinh (\xi _{j}+\eps \lambda_{i}-\eta )}
     \bigg\}
         \notag \\
         &\times
  \prod\limits_{i\in \alpha _{+}}\bigg\{
  \prod\limits_{j\in \gamma_{+}}
  \frac{\sinh (\xi _{j}-\lambda _{i}^{\sigma })}
       {\sinh (\xi_{j}-\lambda _{i}^{\sigma }-\eta )}\
  \frac{\prod\limits_{j\in\alpha _{+}}\sinh (\lambda _{ji}^{\sigma }+\eta )}
       {\prod\limits_{j\in \alpha_{+}-\{i\}}\sinh (\lambda _{ji}^{\sigma })}
     \bigg\}
           \nonumber\\
       &\times
   \prod\limits_{p=1}^{s}
   \frac{\prod\limits_{k=i_{p}+1}^{m}\sinh (\lambda_{b_{p}}^{\sigma }-\xi _{k}+\eta )}
        {\prod\limits_{k=i_{p}}^{m}\sinh (\lambda_{b_{p}}^{\sigma }-\xi _{k})}
   \prod\limits_{p=s+1}^{m}
   \frac{\prod\limits_{k=i_{p}+1}^{m}\sinh (\xi _{k}-\lambda _{b_{p}}^{\sigma}+\eta )}
        {\prod\limits_{\substack{ k=i_{p}  \\ \negspace k\neq N+m+1-b_{p}  \negspace}}^{m}  \negspace\!\!
     \sinh(\xi _{k}-\lambda _{b_{p}}^{\sigma })}.
\end{align}
Here, the sum is performed over all $\sigma_j\in\{+,-\}$ for $j\in\alpha_+$, we have defined
$\lambda _{i}^{\sigma }:=\sigma _{i}\lambda _{i}$ for
$i\in \beta_m$, with
$\sigma _{i}=1$ if $i>N$,
and
\begin{alignat*}{2}
  &\alpha _{+}=\beta _{m}\cap \{1,\ldots,N\}, &
                 &\alpha _{-}=\{1,\ldots,N\}\setminus\alpha_{+},\\
  &\gamma _{-}=\{N+m+1-j\}_{ j\in \beta _{m}\cap\{N+1,\ldots,N+m\}},\quad  &
                 &\gamma _{+}=\{1,\ldots,m\}\setminus\gamma _{-}.
\end{alignat*}
The function $H_{\sigma }^{\mathcal{B}_{+}}(\{\lambda \})$ is the coefficient \eqref{HB+} appearing in the
boundary-bulk decomposition.
\end{prop}

\section{Elementary building blocks of correlation functions}
\label{sec-blocs}

\subsection{Finite chain}

It is now a matter of straightforward calculations to derive the
expression of elementary building blocks of correlation functions at
zero temperature. They are given as the ground state average value
of products of elementary operators of the form
\begin{equation}
   \bracket{\prod_{j=1}^m E_j^{\eps_j,\eps'_j}}
   =\frac{\bra{\psi_+(\{\la\})}E_1^{\eps_1,\eps'_1}\ldots E_m^{\eps_m,\eps'_m}\ket{\psi_+(\{\la\})}}
         {\bracket{\psi_+(\{\la\})\mid\psi_+(\{\la\})}}
   ,
\end{equation}
in which $\la_1,\ldots,\la_N$ denote the ground state rapidities\footnote{Note that, due to
Proposition~\ref{prop 5}, we could have also chosen to compute this average value
by means of states - instead of states +. }.

Using Proposition~\ref{prop-act-bound} and the partial scalar product expression of
Section~\ref{sec-part-sc}, we obtain for the finite chain:

\begin{prop}\label{prop-blocs}
The boundary elementary building blocks  can be written as
\begin{equation}\label{el-bloc-fini}
  \bracket{\prod_{j=1}^m E_j^{\eps_j,\eps'_j}}
   =\sum_{b_1=1}^N\ldots\sum_{b_s=1}^N\sum_{b_{s+1}=1}^{N+m}\ldots\sum_{b_m=1}^{N+m}
    \frac{H_{\{b_j\}}^{+}(\{\lambda \})}
         {\prod\limits_{1\leq l<k\leq m}\!\!\!\!\sinh \xi _{kl}
             \prod\limits_{1\leq p\leq q\leq m}\!\!\!\!\sinh (\bar{\xi}_{pq}-\eta )},
\end{equation}
in which
\begin{align}
   H_{\{b_j\}}^{+}(\{\lambda  \})
    =& \sum\limits_{\sigma _{b_j}}
    \frac{(-1)^{s'}\prod\limits_{i=1}^{m}\sigma_{b_i}
          \prod\limits_{i=1}^{m}\prod\limits_{j=1}^{m}
      \sinh (\lambda_{b_{i}}^{\sigma }+\xi _{j}-\eta )}
     {\prod\limits_{1\leq i<j\leq m}\sinh(\lambda _{b_{i}b_{j}}^{\sigma }+\eta )
                          \sinh (\bar{\lambda}_{b_{i}b_{j}}^{\sigma }-\eta )}
    \prod\limits_{k=1}^{m}\frac{\sinh (\xi _{k}+\xi _{-}-\eta /2)}
                               {\sinh (\lambda _{b_{k}}^{\sigma}+\xi _{-}-\eta /2)}
                   \notag \\
     & \times
    \prod\limits_{p=1}^{s}\bigg\{
          \prod\limits_{k=1}^{i_{p}-1}\sinh(\lambda _{b_{p}}^{\sigma }-\xi _{k})
      \prod\limits_{k=i_{p}+1}^{m}\!\!\sinh(\lambda _{b_{p}}^{\sigma }-\xi _{k}+\eta )
                  \bigg\}
                        \notag\\
     & \times \!\!
    \prod\limits_{p=s+1}^{m}\bigg\{
          \prod\limits_{k=1}^{i_{p}-1}\sinh(\lambda _{b_{p}}^{\sigma }-\xi _{k})
      \prod\limits_{k=i_{p}+1}^{m}\!\!\sinh(\lambda _{b_{p}}^{\sigma }-\xi _{k}-\eta )
                   \bigg\} \  \det_m\Omega.
\end{align}
In this expression, the sum is performed over all $\sigma_{b_j}\in\{+,-\}$ for $b_j\le N$,
and $\sigma_{b_j}=1$ for $b_j> N$,
and the $m\times m$ matrix $\Omega $ is given in terms of the matrix
$\mathcal{S}$ of Section~\ref{sec-part-sc} as
\begin{alignat}{2}
&\Omega _{lk}=-\delta _{N+m+1-b_{l},k}, \quad & &\text{for }b_{l}>N, \\
&\Omega _{lk}=\mathcal{S}_{b_{l},k}, \quad & &\text{for }b_{l}\leq N.
\end{alignat}
\end{prop}

\subsection{Half-infinite chain}

Let us now consider the thermodynamic limit $M\tend\infty$ of this quantity.

In the case where  all the roots $\alpha_j$ describing
the ground state are real, i.e. if $\tilde\xi_->\zeta/2$ or $\tilde\xi_-<0$ (see Section~\ref{sec-gs}),
the sums over the indices $b_j$ from $1$ to $N$ become, as in the bulk case,
integrals over the density of the ground state. More precisely,
we have to perform the replacement
\begin{equation*}
\frac{1}{M}\sum_{b_j=1}^N \sum_{\sigma_{b_j=\pm}} \sg_{b_j} f(\la_{b_j}^\sigma)
\underset{N\tend\infty}{\ttend}
\int\limits_0^{\Lambda}\dd\la_j\,\rho(\la_j)\sum_{\sg_j={\pm}} \sg_{j} \,f(\la_j^\sg)
= \int\limits_{-\Lambda}^{\Lambda}\dd\la_j\, f(\la_j)\,\rho(\la_j).
\end{equation*}
Moreover, the  sums over $b_j>N$ can be written as contour integrals thanks
to the identity
\begin{equation}
    2i\pi\, \text{Res}\rho(\la,\xi)\pour{\la=\xi}=-2.
\end{equation}
In the region $0<2\tilde\xi_-<\zeta$, one should also take into account the existence of the complex root. The
term of the sum which corresponds to $\check\la$ can also be written as a contour integral since
\begin{equation}
    \text{Res}\Big[\frac{1}{\sh(\la+\xi_--\eta/2)}\Big]\pour{\la=\check\la}=1.
\end{equation}

Therefore, one obtains:
\begin{multline}
 \bracket{\prod_{j=1}^m E_j^{\eps_j,\eps'_j}}=
   \frac{(-1)^{m-s}}
        {\pl_{j<i}\sh(\xi_{ij})\pl_{i\le j}\sh(\overline{\xi}_{ij}-\eta)}
             \\
   \times
      \int\limits_{\mathcal{C}} \prod_{j=1}^s \dd \la_j  \
      \int\limits_{\mathcal{\tilde{C}}}\prod_{j=s+1}^m \!\! \dd \la_{j}\
           H_m(\paa{\la_j};\paa{\xi_k})\
     \det_m\pac{\Phi\pa{\la_j,\xi_k}},
\end{multline}
with
\begin{equation}
   \Phi\pa{\la_j,\xi_k}
   =\frac{1}{2}\big[\rho(\la_j,\xi_k)-\rho(\la_j,\eta-\xi_k)\big],
\end{equation}
and
\begin{align}
    H_m(\paa{\la_j};\paa{\xi_k})
    &=\frac{\prod\limits_{j=1}^{m}\prod\limits_{k=1}^{m}
                   \sinh (\lambda_j +\xi_k-\eta )}
     {\negspace \prod\limits_{1\leq i<j\leq m}\negspace
            \sinh(\lambda _{ij}+\eta )\,\sinh (\bar{\lambda}_{ij}-\eta )}\
    \prod\limits_{k=1}^{m}\frac{\sinh (\xi _{k}+\xi _{-}-\eta /2)}
                               {\sinh (\lambda _{k}+\xi _{-}-\eta /2)}
                   \notag \\
     & \quad\times
    \prod\limits_{p=1}^{s}\bigg\{
          \prod\limits_{k=1}^{i_{p}-1}\sinh(\lambda _{p}-\xi _{k})
      \prod\limits_{k=i_{p}+1}^{m}\!\!\sinh(\lambda _{p}-\xi _{k}+\eta )
                  \bigg\}
                        \notag\\
     & \quad\times \!\!
    \prod\limits_{p=s+1}^{m}\bigg\{
          \prod\limits_{k=1}^{i_{p}-1}\sinh(\lambda _{p}-\xi _{k})
      \prod\limits_{k=i_{p}+1}^{m}\!\!\sinh(\lambda _p-\xi _{k}-\eta )
                   \bigg\}.
\end{align}
The contours $\mathcal{C}$ and $\mathcal{\tilde{C}}$ depend on the boundary field. They are defined as
\begin{align}
    &\mathcal{C}=\begin{cases}
       [-\Lambda,\Lambda]\cup\Gamma(\check\la)
           &\text{if $0<\tilde\xi_-<\zeta/2$},\\
       [-\Lambda,\Lambda]
           &\text{otherwise},
                   \end{cases}  \\
    &\mathcal{\tilde{C}}=\mathcal{C}\cup\Gamma(\{\xi_k\}).
\end{align}
where $\Gamma(\check\la)$ (respectively $\Gamma(\{\xi_k\})$) surrounds $\check\la$ (respectively $\xi_1,\ldots,\xi_m$)
with index 1, all other poles being outside.

\begin{rem}
  One can easily verify, as a consistency check for the above formula, that the reduction property
\begin{equation*}
  \bracket{E_1^{\eps_1,\eps'_1}\ldots E_m^{\eps_m,\eps'_m} E_{m+1}^{1,1}}
+\bracket{E_1^{\eps_1,\eps'_1}\ldots E_m^{\eps_m,\eps'_m} E_{m+1}^{2,2}}
=\bracket{E_1^{\eps_1,\eps'_1}\ldots E_m^{\eps_m,\eps'_m}}
\end{equation*}
from $m+1$ sites to $m$ sites is satisfied.
\end{rem}

\bigskip

Let us finally rewrite explicitly this result in the two different
regimes (massive and massless) of the $XXZ$ model, using the fact
that, in both regimes, the determinant of the matrix $\Phi$ can be
calculated explicitly (the corresponding expressions are given in
Appendix~\ref{append-det}).

In the massless case, one gets directly:
\begin{align}
  \bracket{\prod_{j=1}^{m} E_{j}^{\varepsilon_{j},\varepsilon_{j}^{\prime}}}
   & =
   \frac{\prod\limits_{a=1}^{m}\cosh \big(\frac{\pi }{\zeta }\xi_{a}\big)
         \prod\limits_{k<l}
                \big[\sinh \big(\frac{\pi }{\zeta }\xi _{kl}\big)\,
                     \sinh \big(\frac{\pi }{\zeta }\bar{\xi}_{kl}\big)  \big]}
        {\prod\limits_{j<i}\sinh (\xi_{ij})
         \prod\limits_{i\leq j}\sinh (\overline{\xi }_{ij}+i\zeta )}
     \int\limits_{\mathcal{C}}
     \prod\limits_{j=1}^{s}\Big( i\frac{\dd\lambda _{j}}{\zeta } \Big)
                                        \nonumber\\
    &\ \times
     \int\limits_{\mathcal{\tilde{C}}}
   \prod\limits_{j=s+1}^{m} \Big(
     \frac{\dd\lambda _{j}}{i\zeta }\Big)
     \
   \prod\limits_{a=1}^{m}\prod\limits_{k=1}^{m}
   \frac{\sinh (\lambda _{a}+\xi _{k}+i\zeta )\ }
        {\sinh \frac{\pi }{\zeta }(\lambda _{a}-\xi _{k})\,
         \sinh \frac{\pi }{\zeta }(\lambda_{a}+\xi _{k})}
                                          \notag \\
    &\  \times
   \prod\limits_{k<l}
   \frac{\sinh (\frac{\pi }{\zeta }\lambda_{lk})\,
         \sinh (\frac{\pi }{\zeta }\bar{\lambda}_{lk})}
        {\sinh (\lambda_{kl}-i\zeta )\,\sinh (\bar{\lambda}_{kl}+i\zeta )}
   \prod\limits_{k=1}^{m}%
   \frac{\sinh (\frac{\pi }{\zeta }\lambda _{k})\,
         \sinh (\xi _{k}+i\frac{\zeta }{2}+\xi _{-})}
        {\sinh (\lambda _{k}+i\frac{\zeta }{2}+\xi _{-})}
                                            \notag \\
    &\  \times
   \prod\limits_{p=1}^{s}\bigg\{
      \prod\limits_{k=1}^{i_{p}-1}\sinh(\lambda _{p}-\xi _{k})
      \prod\limits_{k=i_{p}+1}^{m}\!\!\sinh (\lambda_{p}-\xi _{k}-i\zeta )\bigg\}
                                            \notag \\
    &\  \times
   \!\!\prod\limits_{p=s+1}^{m}\bigg\{
      \prod\limits_{k=1}^{i_{p}-1}\sinh (\lambda _{p}-\xi _{k})
      \prod\limits_{k=i_{p}+1}^{m}\!\!\sinh (\lambda_{p}-\xi _{k}+i\zeta )\bigg\},
\end{align}
in which $\mathcal{\widetilde{C}}=\mathcal{C}\cup\Gamma(\{\xi_k\})$, with
\begin{equation}
\mathcal{C}=\begin{cases}
             \Rset
             & \text{for $\tilde\xi_-<0$ or $\tilde\xi_->\zeta/2$}, \\
             \Rset\cup\Gamma\big(-i(\zeta/2+\tilde\xi_-)\big)
             & \text{for $0<\tilde\xi_-<\zeta/2$.}
            \end{cases}
\end{equation}
In the homogeneous limit $\xi _{j}=-i\zeta /2$, these elementary
blocks have the following form:
\begin{align}
  \langle \,\prod_{j=1}^{m}
                 E_{j}^{\varepsilon _{j},\varepsilon _{j}^{\prime}}\,\rangle
  =&
    (-1)^{m-s + \frac{m(m-1)}{2}}\sinh ^{m}\xi _{-} \Big( \frac{\pi }{\zeta }\Big)^{m(m+1)}
    \int\limits_{\mathcal{C}}
    \prod\limits_{j=1}^{s} \frac{\dd\lambda _{j}}{2\zeta }
    \ \cdot\
    \int\limits_{\mathcal{\tilde{C}}}
      \prod\limits_{j=s+1}^{m}\frac{\dd\lambda _{j}}{2\zeta }
                              \notag \\
   &\quad \times
     \prod\limits_{k<l}
     \frac{\sinh \big(\frac{\pi }{\zeta }\lambda_{kl}\big)\,
           \sinh \big(\frac{\pi }{\zeta }\bar{\lambda}_{kl}\big)}
          {\sinh (\lambda_{kl}-i\zeta )\,\sinh (\bar{\lambda}_{kl}+i\zeta )}
     \
     \prod\limits_{k=1}^{m}
     \frac{\sinh \big(\frac{\pi }{\zeta }\lambda _{k}\big)}
          {\sinh (\lambda _{k}+i\frac{\zeta }{2}+\xi _{-})}
                               \notag \\
   &\quad \times
     \prod\limits_{p=1}^{s}
     \frac{\sinh ^{m+i_{p}-1}\big(\lambda _{p}+i\frac{\zeta }{2}\big)\,
           \sinh ^{m-i_{p}}\big(\lambda _{p}-i\frac{\zeta }{2}\big)}
          {\cosh ^{2m}\big(\frac{\pi }{\zeta }\lambda _{p}\big)}
                                \notag \\
   &\quad \times
     \!\!\prod\limits_{p=s+1}^{m}
     \frac{\sinh ^{m+i_{p}-1}\big(\lambda _{p}+i\frac{\zeta }{2}\big)\,
           \sinh^{m-i_{p}}\big(\lambda _{p}+i\frac{3\zeta }{2}\big)}
          {\cosh^{2m}\big(\frac{\pi }{\zeta }\lambda _{p}\big)}.
\end{align}

To obtain the explicit expression of the elementary blocks in the massive regime, one performs the change of
variables $\alpha_j=i\la_j$, $\beta_k=i\xi_k$. Hence, using the corresponding representations for the determinants
of the matrix $\Phi $, one obtains:
\begin{align}
  \langle \,\prod_{j=1}^{m}E_{j}^{\varepsilon _{j},
                       \varepsilon _{j}^{\prime}}\,\rangle
  & =\frac{\prod\limits_{a=1}^{m}
           \big[\theta _{3}(\beta_{a})\,\theta _{4}(\beta _{a})\big]\,
           \prod\limits_{k<l}
           \big[\theta _{1}(\beta_{kl})\,\theta _{1}(\bar{\beta}_{kl})\big]}
          {\prod\limits_{j<i}\sin (\beta_{ij})\,
           \prod\limits_{i\leq j}\sin (\bar{\beta}_{ij}+i\zeta )}
                        \notag \\
 & \times
     \int\limits_{\underline{\mathcal{C}}}
     \prod\limits_{j=1}^{s}\Big(i\frac{\dd\alpha _{j}}{\pi } \Big)
     \int\limits_{\underline{\mathcal{\tilde{C}}}}
     \prod\limits_{j=s+1}^{m}\Big(\frac{\dd\alpha _{j}}{i\pi }\Big)\
     \prod\limits_{a=1}^{m}\prod\limits_{k=1}^{m}
     \frac{\sin (\alpha _{a}+\beta _{k}+i\zeta )}
          {\theta_{1}(\alpha _{a}-\beta _{k})\,\theta _{1}(\alpha _{a}+\beta _{k})}
                         \notag \\
 & \times
     \prod\limits_{k<l}
     \frac{\theta _{1}(\alpha _{lk})\,\theta _{1}(\bar{\alpha}_{lk})}
          {\sin (\alpha _{kl}-i\zeta )\,\sin (\bar{\alpha}_{kl}+i\zeta )}
     \prod\limits_{k=1}^{m}
     \frac{\theta _{1}(\alpha_{k})\,\theta _{2}(\alpha _{k})\,
           \sin \big(\beta _{k}+i\frac{\zeta }{2}+i\xi _{-}\big)}
          {\sin \big(\alpha _{k}+i\frac{\zeta }{2}+i\xi _{-}\big)}
                          \notag \\
 & \times
      \prod\limits_{p=1}^{s}
      \bigg\{\prod\limits_{k=1}^{i_{p}-1}\sin(\alpha _{p}-\beta _{k})
             \prod\limits_{k=i_{p}+1}^{m}\!\!\sin (\alpha_{p}-\beta _{k}-i\zeta )
      \bigg\}
                           \notag \\
 & \times
      \!\!\prod\limits_{p=s+1}^{m}
      \bigg\{\prod\limits_{k=1}^{i_{p}-1}\sin(\alpha _{p}-\beta _{k})
             \prod\limits_{k=i_{p}+1}^{m}\!\!\sin (\alpha_{p}-\beta _{k}+i\zeta )
      \bigg\},
\end{align}
in which $\theta_i(\la)\equiv\theta_i(\la,q)$, with $q=e^{-\zeta}$. The integration contours are
$\underline{\mathcal{\widetilde{C}}}=\underline{\mathcal{C}}\cup\Gamma(\{\beta_k\})$, with
\begin{equation}
\underline{\mathcal{C}}=\begin{cases}
             [-\pi/2,\pi/2]
             & \text{for $\tilde\xi_-<0$ or $\tilde\xi_->\zeta/2$}, \\
             [-\pi/2,\pi/2]\cup\Gamma\big(-i(\zeta/2+\tilde\xi_-)\big)
             & \text{for $0<\tilde\xi_-<\zeta/2$.}
            \end{cases}
\end{equation}
In the homogenous limit $\beta _{j}=-i\zeta /2$, the elementary building blocks for the correlation functions are given as
\begin{align}
 \langle \,\prod_{j=1}^{m}&E_{j}^{\varepsilon _{j},
                      \varepsilon _{j}^{\prime}}\,\rangle
  =2^{m(m+1)}\;\sinh ^{m}\xi _{-}\  q^{-\frac{m(m-1)}{4}}\,
      \prod\limits_{n=1}^{\infty}\big[(1+q^{2n})^{2m}\,(1-q^{2n})^{m(3m+1)}\big]
                           \notag \\
 &\times (-1)^{\frac{m(m-1)}{2}}
      \int\limits_{\underline{\mathcal{C}}}
      \prod\limits_{j=1}^{s}\Big(i\frac{\dd\alpha _{j}}{2\pi }\Big)
      \int\limits_{\underline{\mathcal{\tilde{C}}}}
      \prod\limits_{j=s+1}^{m}\Big(\frac{\dd\alpha _{j}}{2i\pi }\Big)\
      \prod\limits_{k<l}
      \frac{\theta _{1}(\alpha_{kl})\,\theta _{1}(\bar{\alpha}_{kl})}
           {\sin (\alpha _{kl}-i\zeta )\,\sin (\bar{\alpha}_{kl}+i\zeta )}
                           \notag \\
 &\times
      \prod\limits_{k=1}^{m}
      \frac{\theta _{1}(\alpha _{k})\,\theta_{2}(\alpha _{k})}
           {\sin \big(\alpha _{k}+i\frac{\zeta }{2}+i\xi _{-}\big)}\
      \prod\limits_{p=1}^{s}
      \frac{\sin ^{m+i_{p}-1}\big(\alpha _{p}+i\frac{\zeta }{2}\big)\,
            \sin ^{m-i_{p}}\big(\alpha _{p}-i\frac{\zeta }{2}\big)}
           {\theta _{4}^{2m}(\alpha_{p})}
                           \notag \\
 & \times
      \!\!\prod\limits_{p=s+1}^{m}
      \frac{\sin ^{m+i_{p}-1}\big(\alpha _{p}+i\frac{\zeta }{2}\big)\,
            \sin ^{m-i_{p}}\big(\alpha _{p}+i\frac{3\zeta }{2}\big)}
           {\theta_{4}^{2m}(\alpha _{p})}.
\end{align}

Let us finally remark that all these computations can also be performed in the case of an external magnetic
field along the $S^z$ direction. In that case, the integration contours and the density function will depend, like
in the bulk case, on this external magnetic field.


\section*{Acknowledgments}

J.M. M., N. S. and V. T. are supported by CNRS.
N. K., K. K., J.M. M. and V. T. are supported  by the ANR programm GIMP ANR-05-BLAN-0029-01.
N. K., G. N. and V. T. are supported by the ANR programm MIB-05 JC05-52749.
N. S. is supported by the French-Russian Exchange Program, the
Program of RAS Mathematical Methods of the Nonlinear Dynamics, RFBR-05-01-00498, Scientific Schools 672.2006.1.
N. K and N. S. would like to thank the Theoretical Physics group of the Laboratory of Physics at ENS Lyon for hospitality,
which makes this collaboration possible.


\section{Appendices}

\renewcommand{\theequation}{\Alph{subsection}.\arabic{equation}}
\renewcommand{\thesubsection}{\Alph{subsection}}

\subsection{Boundary creation and annihilation operators}
\label{append-F}

Using the quadratic relations (\ref{U-})-(\ref{U+}), one can express the boundary operators
$\cA_\pm,\cB_\pm,\cC_\pm,\cD_\pm$ in terms of the bulk operators.
Note that it may sometimes be more convenient to rewrite the creation and annihilation boundary
operators $\mathcal{B}_{\pm }$ and $\mathcal{C}_{\pm }$ in the form
\begin{align}
  &\mathcal{B}_{-}(\lambda ) = -\gamma(\la)\,
   \frac{\sinh (2\lambda -\eta )}{\sinh(2\lambda )}\,
   \big[ B(-\lambda )\, A(\lambda )\, \sinh (\lambda +\xi _{-}-\eta /2)
              \notag \\
  &\hspace{6cm} +B(\lambda )\,A(-\lambda )\,\sinh (\lambda -\xi _{-}+\eta /2)\big],
\label{Decomp-B-} \\
  &\mathcal{C}_{-}(\lambda ) =\gamma(\la)\,
  \frac{\sinh (2\lambda -\eta )}{\sinh(2\lambda )}\,
  \big[D(-\lambda )\,C(\lambda )\,\sinh (\lambda +\xi _{-}-\eta /2)
               \notag \\
  &\hspace{6cm} +D(\lambda )\,C(-\lambda )\,\sinh (\lambda -\xi _{-}+\eta /2)\big],
\label{Decomp-C-}
\end{align}
and
\begin{align}
  &\mathcal{B}_{+}(\lambda ) =\gamma(\la)\,
  \frac{\sinh (2\lambda +\eta )}{\sinh(2\lambda )}\,
  \big[ B(-\lambda )\,D(\lambda )\,\sinh (\lambda -\xi _{+}+\eta /2)
\notag \\
  & \hspace{6cm}
  +B(\lambda )\,D(-\lambda )\,\sinh (\lambda +\xi _{+}-\eta /2)\big], \\
  &\mathcal{C}_{+}(\lambda ) =-\gamma(\la)\,
  \frac{\sinh (2\lambda +\eta )}{\sinh(2\lambda )}\,
  \big[A(-\lambda )\,C(\lambda )\,\sinh (\lambda -\xi _{+}+\eta /2)
\notag \\
  & \hspace{6cm}
  +A(\lambda )\,C(-\lambda )\,\sinh (\lambda +\xi _{+}-\eta /2)\big].
\label{Decomp-C+}
\end{align}

It is then convenient, for the computation of partition functions
and scalar products, to express the boundary operators in the $F$
and $\overline{F}$-basis. The concept of factorizing $F$-matrices
was defined in \cite{MaiS00}, following the concept of twists
introduced by Drinfel'd in the theory of Quantum Groups
\cite{Dri87}, and we refer to \cite {MaiS00} for the explicit
construction of the $F$ and $\overline{F}$-matrices in the periodic
$XXZ$ spin-1/2 chain and for the representations of the bulk
operators in the $F$ and $\overline{F}$-basis. Using the  $F$ and
$\overline{F}$-basis expression of the  bulk operator, one obtains
the following result:

\begin{lemma}
\label{prop 2}
Let $\mathcal{\widetilde{X}}_{\pm }$ denote the expressions of
the boundary operators $\mathcal{X}_{\pm }$ in the $F$-basis,
and $\mathcal{\overline{X}}_{\pm }$ their  expressions in the
$\overline{F}$-basis. Then,
\begin{align}
  &\mathcal{\widetilde{B}}_{+}(\lambda )
  =-\sum_{i=1}^{M} u(-\lambda ,\xi_{i}|\xi _{+})\
   {\sigma }_{i}^{-}\underset{j\neq i}{\otimes}
   \begin{pmatrix}
b(-\lambda -\xi _{j})\,b(\lambda -\xi _{j}) & 0 \\
0 & b^{-1}(\xi _{ji})
   \end{pmatrix}_{[j]},
      \label{B+Fbase}\\
 &\mathcal{\widetilde{C}}_{-}(\lambda )
 =\sum_{i=1}^{M} u(\lambda ,\xi _{i}|\xi_{-})\
  {\sigma }_{i}^{+}\underset{j\neq i}{\otimes}
   \begin{pmatrix}
b(\lambda -\xi _{j})\,b(-\lambda -\xi _{j})\,b^{-1}(\xi _{i}-\xi _{j}) & 0 \\
0 & 1
   \end{pmatrix}_{[j]},
     \label{C-Fbase}
\end{align}
and
\begin{align}
 &\mathcal{\overline{B}}_{-}(\lambda )
 =-\sum_{i=1}^{M} u(\lambda ,\xi _{i}|-\xi_{-})\
  {\sigma }_{i}^{-}\underset{j\neq i}{\otimes}
   \begin{pmatrix}
1 & 0 \\
0 & b(-\lambda -\xi _{j})\,b(\lambda -\xi _{j})\,b^{-1}(\xi _{ij})
   \end{pmatrix}_{[j]},
     \label{B-Fbasis}\\
 &\mathcal{\overline{C}}_{+}(\lambda )
 =\sum_{i=1}^{M} u(-\lambda ,\xi _{i}|-\xi_{+})\
 {\sigma }_{i}^{+} \underset{j\neq i}{\otimes}
   \begin{pmatrix}
b^{-1}(\xi _{ji}) & 0 \\
0 & b(\lambda -\xi _{j})\,b(-\lambda -\xi _{j})
   \end{pmatrix}_{[j]},
      \label{C+Fbasis}
\end{align}
where
\begin{equation}\label{expr-u}
  u(\lambda ,\xi |x)= \gamma(\la)\, a(\lambda )\,a(-\lambda )\,
  \frac{\sinh \eta \, \sinh(2\lambda -\eta ) \, \sinh (x+\xi -\eta /2)}
       {\sinh (\lambda -\xi +\eta ) \, \sinh(\lambda +\xi -\eta )}.
\end{equation}
\end{lemma}

\subsection{Partition function}\label{append-part}

In this appendix, we propose a proof of Proposition~\ref{prop 7}.
Similarly as in \cite{KitMT99}, this derivation is based on direct
calculations in the basis induced by the twist $F$ introduced in
\cite{MaiS00}. Indeed, in this particular basis (called $F$-basis),
the explicit expressions of the bulk operators $A$, $B$, $C$, $D$
simplify drastically. Since the boundary creation and annihilation
operators are quadratic in terms of the bulk operators, they have
themselves much simpler expressions in this $F$-basis (see
Appendix~\ref{append-F} for details). As moreover the states
$\bra{0} $ and $\ket{\overline{0}}$ are respectively invariant under
the left-action of $F$ and the right-action of $F^{-1}$, the
partition function can be directly written in the $F$-basis as
\begin{equation}
\mathcal{Z}_{M}^{\mathcal{C}_{-}}(\{\lambda _{\alpha }\},\{\xi_{k}\};\xi_-)
   =\bra{0} \,\mathcal{\widetilde{C}}_{-}(\lambda_{1})\ldots \mathcal{\widetilde{C}}_{-}(\lambda _{M})\,
    \ket{\overline{0}},
\end{equation}
where $\mathcal{\widetilde{C}}_{-}(\la)=F\,\mathcal{C}_{-}(\la)\
F^{-1}$ is the expression of $\mathcal{C}_{-}(\la)$ in the basis induced by $F$.
Using now the expression (\ref{C-Fbase}) of the
operator $\mathcal{\widetilde{C}}_{-}(\la)$, one obtains a new recursion formula
for the partition function, which corresponds to a development of the
determinant in~(\ref{Partition-C-}).
Indeed, acting with $\mathcal{\widetilde{C}}_{-}(\lambda _{M})$ on the state $\ket{\overline{0}}$, one has
\begin{equation}
  \mathcal{Z}_{M}^{\mathcal{C}_{-}}
          ( \{\lambda _{\alpha }\} , \{\xi_{j}\} ; \xi_- )
  =\sum_{i=1}^{M} u(\lambda _{M},\xi_{i}|\xi _-)\,
    \bra{0} \,\mathcal{\widetilde{C}}_{-}(\lambda_{1})\ldots
    \mathcal{\widetilde{C}}_{-}(\lambda _{M-1})\,\ket{\overline{i}},
\end{equation}
where $\ket{\overline{i}}$ is the vector with all spins down except
in site $i$ and where the expression of $u(\lambda _{M},\xi_{i}|\xi _-)$ is given by formula~\eqref{expr-u}.
Since $(\sigma _{i}^{+})^{2}=0$, the action of the other operators
$\mathcal{\widetilde{C}}_{-}(\lambda _{\alpha })$, $1\leq \alpha \leq M-1$,
on the vector $\ket{\overline{i}}$ is diagonal on the space $i$, so that we
obtain the following recursion formula for
$\mathcal{Z}_{M}^{\mathcal{C}_{-}} $:
\begin{multline}
  \mathcal{Z}_{M}^{\mathcal{C}_{-}}
  ( \{\lambda _{\alpha }\} , \{\xi_{j}\} ; \xi_- )
  =\sum_{i=1}^{M} c_{M}(\lambda_{M} , \xi _{i} , \{\xi _{j}\} ;\xi_-)\\
  \times
  \mathcal{Z}_{M-1}^{\mathcal{C}_{-}}
  ( \{\lambda_{\alpha }\}_{\alpha \not=M},\{\xi_{j}\}_{j\not=i};\xi_-) .
   \label{recZ}
\end{multline}
The coefficient of the recursion is
\begin{equation}
c_{M}(\lambda _{M},\xi _{i},\{\xi _{j}\};\xi_-)
  =u(\lambda _{M},\xi _{i}|\xi_{-})
  \prod_{k=1}^{M-1}\!\! \big[ b(\lambda_k-\xi_{i})\, b(-\lambda_k-\xi_i)\big]\,
            \prod_{j\ne i} b^{-1}(\xi_{ji}),\!\!
\end{equation}
which, as a meromorphic function of $\la_M$, can be rewritten as
\begin{multline}
c_{M}(\lambda _{M},\xi _{i},\{\xi _{j}\};\xi_-)
=\gamma(\la_M)\, a(\la_M)\,a(-\la_M)\,\sinh (\lambda_M-\xi_i)\, \sinh (\lambda_M+\xi_i)\\
 \hspace{1.5cm}\times
\overset{M-1}{\underset{\beta =1}{\prod }}
 \frac{\sinh (\lambda _{\beta }-\xi _{i})\sinh (\lambda _{\beta }+\xi _{i})}
      {\sinh \lambda _{M\beta }\sinh \overline{\lambda }_{M\beta}}\,
 \overset{M}{\underset{\substack{j=1 \\ j\neq i}}{\prod }}
 \frac{\sinh(\lambda _{M}-\xi _{j})\sinh (\lambda _{M}+\xi _{j})}
      {\sinh \xi_{ji}\sinh (\overline{\xi }_{ji}-\eta )}  \\
\times \bigg\{\mathcal{N}_{M,i}^{\mathcal{C}_{-}}(\lambda _{M},\xi _{i};\xi_-)
         -\sum_{\beta =1}^{M-1} g_{\beta }\,
       \mathcal{N}_{\beta ,i}^{\mathcal{C}_{-}}(\lambda _{\beta },\xi _{i};\xi_-)
       \bigg\} .
\end{multline}
with
\begin{multline*}
 g_{\beta }=\frac{1}{\sinh 2\lambda _{\beta }\,\sinh (2\lambda _{\beta }-\eta )}
\biggr( \frac{\sinh (2\lambda _{\beta }+\eta )}
            {\sinh \overline{\lambda }_{M\beta }}
      +\frac{\sinh (2\lambda _{\beta }-\eta )}{\sinh \lambda _{M\beta }} \biggl) \\
\times
  \frac{\overset{M-1}{\underset{k=1}{\prod }}
  \big[ \sinh \lambda _{Mk}\, \sinh \overline{\lambda }_{Mk}\big]}
   {\overset{M-1}{\underset{\underset{k\neq \beta }{k=1}}{\prod }}
  \big[ \sinh \lambda _{\beta k}\,\sinh \overline{\lambda }_{\beta k} \big] }\,
\overset{M}{\underset{j=1}{\prod }}
 \frac{\sinh (\lambda_{\beta }-\xi_{j})\,\sinh (\lambda_{\beta }+\xi _{j})}
      {\sinh (\lambda_{M}-\xi_{j})\,\sinh (\lambda_{M}+\xi _{j})}.
\end{multline*}
This actually corresponds to the development with respect to the last line of the determinant
\begin{multline}
\mathcal{Z}_{M}^{\mathcal{C}_{-}}( \{\lambda_{\alpha }\},\{\xi_{j}\};\xi_-)
   = \pl_{\beta=1}^M \big[\gamma(\la_\beta)\, a(\la_\beta)\, a(-\la_\beta)\big]\\
  \times\frac{\overset{M}{\underset{\beta =1}{\prod }}
                   \overset{M}{\underset{k=1}{\prod }}
   \big[\sinh (\lambda_{\beta }-\xi_{k})\, \sinh(\lambda_{\beta }+\xi_{k})\big]}
  {\underset{\beta <\gamma }{\prod }\big[\sinh\lambda_{\beta \gamma }
                      \sinh \overline{\lambda }_{\beta \gamma }\big]
   \underset{r<s}{\prod }\big[\sinh \xi _{sr}
               \sinh (\overline{\xi }_{sr}-\eta )\big]}\
           \underset{M}\det\mathcal{\hat{N}}^{\mathcal{C}_{-}},
\end{multline}
where $\mathcal{\hat{N}}^{\mathcal{C}_{-}}$ is the matrix obtained from $%
\mathcal{N}^{\mathcal{C}_{-}}$ by substracting to the last line $L_{M}$ the linear
combination of the other lines $\sum_{\beta =1}^{M-1}g_{\beta }L_{\beta }$.
Thus, as $\mathcal{N}^{\mathcal{C}_{-}}$ and $\mathcal{\hat{N}}^{\mathcal{C}%
_{-}}$ have the same determinant, this concludes the proof.

\subsection{Determinant of the densities}
\label{append-det}

In this Appendix, we give the explicit expression of the determinant of the matrix $\Phi$ involving the density function
of the ground state.

In the massless case it is:
\begin{multline}\label{det-massless}
  \det\big[\Phi(\la_j,\xi_k)\big]
      =\Big( \frac{i}{\zeta }\Big)^{m}
       \prod\limits_{a=1}^{m}
         \Big[\sinh \Big(\frac{\pi }{\zeta }\lambda _{a}\Big)\,
              \cosh \Big(\frac{\pi }{\zeta }\xi _{a}\Big)\Big]\\
  \times
   \frac{\prod\limits_{k<l}\big[
         \sinh \big(\frac{\pi }{\zeta }\xi _{kl}\big)\,
         \sinh \big(\frac{\pi }{\zeta }\bar{\xi}_{kl}\big)\,
         \sinh \big(\frac{\pi }{\zeta }\lambda _{lk}\big)\,
         \sinh \big(\frac{\pi }{\zeta }\bar{\lambda}_{lk}\big) \big]}
        {\prod\limits_{a=1}^{m}\prod\limits_{k=1}^{m}\big[
         \sinh \frac{\pi }{\zeta }(\lambda _{a}-\xi _{k})\,
         \sinh \frac{\pi }{\zeta }(\lambda _{a}+\xi _{k})\big]},
\end{multline}

Let us now compute the determinant of the densities in the massive case,
where the density of Bethe roots can be written in terms of theta
functions:
\begin{equation}
   \rho(\la,\xi)=-\f{1}{\pi}
   \pl_{n\geq1}\paf{1-q^{2n}}{1+q^{2n}}^2
   \f{\theta_2\big(i(\la-\xi),q\big)}{\theta_1\big(i(\la-\xi),q\big)},
\end{equation}
with $q=e^\eta=e^{-\zeta}$. We therefore have to compute the following determinant:
\begin{equation}
   \det_m\big[\Phi(\la_j,\xi_k)\big]
   =\paf{-1}{2\pi}^m
   \pl_{n\geq1} \paf{1-q^{2n}}{1+q^{2n}}^{2m}
   \det_m\pac{\f{\theta_2(\a_j-\beta_k)}{\theta_1(\a_j-\beta_k)}  +
              \f{\theta_2(\a_j+\beta_k)}{\theta_1(\a_j+\beta_k)}}
\end{equation}
with $\alpha_j=i\beta_j$, $\beta_k=i\xi_k$.

Let us consider $\det_m\big[\Phi(\la_j,\xi_k)\big]$
as a certain function $f$ of the variable $\alpha_1$. It is an elliptic
function of order $4m$ with periods $\pi$ and $2 i \zeta$. An
irreducible set of poles is
\begin{equation}
\paa{\pm\beta_1,\dots,\pm\beta_m,\pm\beta_1+i\zeta,\dots,\pm\beta_m+i\zeta},
\end{equation}
whereas
\begin{equation}
   \pm\a_2,\dots,\pm\a_m,\pm\a_2+i\zeta,\dots,\pm\a_m+i\zeta
\end{equation}
are zeros of $f$. $f$ being an odd function, $\la=0$ is also a zero and,
since $f\pa{\la}=-f\pa{\la+i\zeta}$, so is $\la=i\zeta$.
Up to congruence, there remain two other zeros which differ by $i\zeta$,
say $x_0$ and $x_0+i\zeta$.
Since the sum of the zeros is congruent to the sum of the poles,
$x_0$ is either congruent to $0$ or to $\tf{\pi}{2}$. In fact, the only
choice compatible with the periods of $f$ is $x_0=\tf{\pi}{2}$. This
means that, up to a constant independent of $\a_1$, $f$ can be factorized as
\begin{equation}
   \theta_1(\a_1)\,\theta_2(\a_1)\,
        \f{\pl_{i=2}^{m}\big[\theta_1(\a_{1i})\,\theta_1({\bar{\a}_{1i}})\big]}
          {\pl_{i=1}^m\big[\theta_1(\a_1-\beta_i)\,\theta_1(\a_1+\beta_i)\big]}.
\end{equation}

This argument can be easily extended to all $\a_j$, $1\le j\le m$,
thanks to the antisymmetry in these variables.
We can also apply a similar procedure to the variables $\beta_k$, the difference being
that we now deal with an even function and that the extra zeros are
$-i\tf{\zeta}{2}$ and $i\tf{\zeta}{2}-\tf{\pi}{2}$.
Finally we obtain
\begin{multline}\label{det-massive}
   \det_m\big[\Phi(\la_j,\xi_k)\big]
    = \Big(-\frac{1}{\pi}\Big)^m \,\pl_{i=1}^m
       \big[\theta_1(\a_i)\,\theta_2(\a_i)\,
            \theta_3(\beta_i)\,\theta_4(\beta_i)\big]
         \\
        \times\f{\pl_{i<j}\big[\theta_1(\a_{ij})\,\theta_1(\bar{\a}_{ij})\,
                      \theta_1(\beta_{ji})\,\theta_1(\bar{\beta}_{ji})\big]}
              {\pl_{i,j=1}^m\big[\theta_1(\a_i-\beta_j)\,\theta_1(\a_i+\beta_j)
                                                                          \big]}.
\end{multline}

\bibliographystyle{/home/vero/LPM/tex/TeX/style-biblio/h-physrev}


\end{document}